\documentstyle[rmp,aps]{revtex}
\begin{document}
\newcommand{\cprl}[3]{#3, Phys. Rev. Lett. {\bf #1}, #2}
\newcommand{\euphl}[3]{#3, Europhys. Lett. {\bf #1}, #2}
\newcommand{\cpra}[3]{#3, Phys. Rev. A {\bf  #1}, #2}
\newcommand{\cprb}[3]{#3, Phys. Rev. B {\bf  #1}, #2}
\newcommand{\cpre}[3]{#3, Phys. Rev. E {\bf  #1}, #2}
\newcommand{\jpa}[3]{#3, J. Phys. A {\bf  #1}, #2}
\newcommand{\jpc}[3]{#3, J. Phys. C {\bf  #1}, #2}
\newcommand{\jpq}[3]{#3, J. Physique {\bf#1}, #2}
\newcommand{\jpI}[3]{#3, J. Physique I {\bf #1}, #2}
\newcommand{\jsp}[3]{#3, J. Stat. Phys. {\bf #1}, #2}
\newcommand{\bl}{\mathop \Delta \limits_\sim}
\newcommand{\avg}[1]{\langle{#1}\rangle}
\newcommand{\U}[1]{\underline{#1}}
\newcommand{\unabla}{\U\nabla}
\newcommand{\req}[1]{(\ref{#1})}
\newcommand{\redue}[2]{(\ref{#1},\ref{#2})}
\newcommand{\beq}{\begin{equation}}
\newcommand{\beqar}{\begin{eqnarray}}
\newcommand{\eeqar}{\end{eqnarray}}
\newcommand{\beqars}{\begin{eqnarray*}}
\newcommand{\eeqars}{\end{eqnarray*}}
\newcommand{\eeq}{\end{equation}}
%\draft

\title{Stochastic Growth Equations and Reparametrization 
Invariance}

\author{Matteo Marsili}
\address{Department of Physics, Shuster Laboratory, 
The University of Manchester, M13 9PL Manchester, UK} 
\author{Amos Maritan}
\address{Istituto Nazionale Fisica della Materia and} 
\address{International School for Advanced Studies (S.I.S.S.A.)
via Beirut 2-4, Trieste 34014 Italy } 
\author{Flavio Toigo}
\address{Istituto Nazionale Fisica della Materia and} 
\address{Dipartimento di Fisica, Padova University, 
via Marzolo 8, 35100 Padova, Italy}
\author{Jayanth R. Banavar}
\address{Department of Physics and Center for Material
Physics, 104 Davey Laboratory,}
\address{The Pennsylvania State University,
University Park, PA 16802, USA}

%\date{\today }
\widetext

\maketitle

\begin{abstract}

It is shown that, by imposing reparametrization invariance, 
one may derive a variety
of stochastic equations describing the dynamics of surface growth 
and identify the physical processes responsible for the various terms. 
This approach provides a particularly transparent way to obtain
continuum growth equations for interfaces. 
It is straightforward to derive equations which describe
the coarse grained evolution of discrete lattice models and 
analyze their small gradient expansion. 
In this way, the authors identify the basic
mechanisms which lead to the most commonly used growth equations.
The advantages of this formulation of growth processes is
that it allows one to go beyond the frequently used no-overhang 
approximation. The reparametrization invariant form 
also displays explicitly the conservation laws for the specific
process and all the symmetries with respect
to space-time transformations which are usually lost in the small
gradient expansion. Finally, it is observed, that the
knowledge of the full equation of motion, beyond the lowest
order gradient expansion, might be relevant in problems
where the usual perturbative renormalization methods fail.

\bgroup
\egroup
\end{abstract}
%\maketitle]
\widetext

\addtolength{\baselineskip}{\baselineskip}

\newpage

\tableofcontents

\section{Introduction}
\label{one}

There has been a great deal of recent work on the formation, growth
and geometry of interfaces (Family and Vicsek, 1991; Meakin, 1993;
Barab\'asi and Stanley, 1995; Halpin-Healy and Zhang, 1995). These studies are 
relevant to a 
variety of experimental situations including biological growth (Eden, 1958), 
the propagation of flame fronts (Sivashinsky, 1977, 1979),
fluid flow in porous media 
(Cieplak and Robbins, 1988; Martys {\it et al.}, 1991) and atomic deposition 
processes (Tu and  Harris,1991; Messier and Yehoda, 1985; 
Bales {\it et al.} 1990; Tang {\it et al.} 1990) such as the 
technologically important 
molecular beam epitaxy (MBE). On a more fundamental level, some 
of these processes are prototype of far--from--equilibrium 
physics without a Hamiltonian formulation (Hohenberg and Halperin, 1977). 
Recent advances have 
shown that it is nevertheless possible and useful to categorize 
the systems into universality classes. There have been two 
principal approaches for the theoretical analysis of such 
problems. The first (see, e.g., Meakin, 1993)
is based on computer simulations of discrete 
models and often it provides useful links between analytic theory 
and experiments. The second approach (Krug and Spohn, 1990;
Kardar, 1994; Halpin-Healy and Zhang, 1995) is to describe the dynamical 
process by stochastic differential equations. This procedure 
neglects the short length--scale details but provides a coarse 
grained description of the interface which is suitable for 
characterizing the asymptotic scaling behavior. There are two 
essential steps to be carried out: first, one must deduce the  continuum 
equation, and second, in  order to obtain the scaling behavior
the equation needs to be solved or analyzed by
renormalization group techniques (Ma and Mazenko, 1975; 
Forster {\it et al.}, 1977; Wolf, 1991; Frei and T\"auber 1994; 
Sun and Plischke, 1994). The principal theme of this 
paper is the derivation of the stochastic partial differential 
equation appropriate to the different physical processes responsible
for the growth. 
We will show that the principle of reparametrization invariance (R--Invariance) 
can be used in a straightforward, yet powerful, manner to derive 
the continuum equation.  This approach may also be useful
in other contexts such as the study of avalanche dynamics associated
with an interface moving in a random medium (Makse, 1995) and 
for the study of interfaces in complex dynamical systems (Kapral {\it et al.},
1994).

We restrict ourselves to local growth processes in which the 
growth rate is a function of the local properties of the 
interface. (Non-local effects are discussed in the last section in terms
of partial differential equations for two coupled fields.) 
Traditionally, continuum equations have been 
derived in the no--overhang approximation (Monge representation). 
The height of the interface on a reference substrate plane,
$h(\vec{r},t)$, is assumed to be a single valued function of the 
lateral coordinates $\vec{r}$ and time $t$. While this 
assumption is valid and yields the correct scaling behavior in 
many instances, it has been demonstrated that some essential 
details are left out in this approach (Keblinski {\it et al.},1994, 1995, 1996).
For example, in the 
columnar growth regime of sputter deposition of thin films, the 
merging and regeneration of columns arises due to the presence of 
outward flaring columns. This phenomenon, observed even  
with a normal incidence of the depositing atoms,  
is not captured within the no--overhang approximation. The R--invariance 
principle will allow us to derive the continuum growth equations 
without such restrictions -- the no--overhang situation can then 
be obtained as a special case. 

Previously, continuum growth equations have been derived 
either directly from  discrete models by Vvedensky {\it et al.}, (1993)
or by using methods based on 
preserving the symmetries and conservation laws of the system
(see, e.g., Hwa and Kardar, 1989; Kardar, 1994). 
The latter approach has been widely used to identify 
universality classes of discrete atomistic models (Liu and Plischke, 1988;
Huse  {\it et al.}, 1990; R\'acz {\it et al.}, 1991;
Das Sarma and Tamborenea, 1991; Das Sarma and Ghaisas, 1992; 
Kotrla {\it et al.}, 1992; Amar and Family, 1993;
Krug  {\it et al.}, 1993).
While this method often unambiguously yield the lowest order 
terms in a gradient expansion that could be present in the 
stochastic differential equation, physical considerations have to 
be invoked to decide whether such terms are indeed present or not. 
In the next section we discuss previous work on the derivation of 
the dynamical equations. Section \ref{three} describes how the 
R--invariance
principle can be used to obtain the growth equations in a variety 
of physical situations. The last section summarizes our results.

The scope of our article is limited -- the emphasis is on theory 
and exclusively
on the derivation of
continuum growth equations and discussions regarding their geometrical and 
physical content. Here and there, we shall mention results which
have been obtained elsewhere for the solutions of some 
of these equations
and their relationship  to physical processes such as  MBE or discrete 
models. Our  discussion will certainly fail to be complete, a
task which goes far beyond our goal of providing a self-contained 
exposition of the derivation of growth equations.   We apologize to
the many authors whose work is not appropriately mentioned in this article.

A quite complete survey of this field can be found in 
several review articles, each with its own focus.
Among the early reviews,   the book 
by Family and Vicsek (1991) and the contributions
by Meakin (1988a) and Krug and Spohn (1990)
have  become standard references.
Specifically,  the first contains a collection
of important  reprints with a commentary,
the second focuses mostly on the phenomenology of simple models of 
growth while the third contains a discussion of analytical approaches.
Another early review is the book by Pietronero and Tosatti
(1985) which describes fractal growth processes.  The effects of a random 
substrate on wetting phenomena have been considered among others by 
Pfeifer {\it et al.}, (1989) and Giugliarelli and Stella (1991, 1994).
Recent work by Halpin-Healy and Zhang (1995)
concentrates  on the Kardar-Parisi-Zhang (1986)
equation and reviews the efforts of the last decade
devoted to elucidating its scaling behavior. From this
central theme,  it extends to cover the field in a pedagogical 
manner, with  special emphasis on directed polymers in random media. 
Numerical simulations of discrete growth models are nicely
reviewed in the work by Meakin (1993).
A quite complete elementary book with an extensive list of references 
by Barab\'asi and Stanley (1995) has recently been published,
where a lucid discussion of
interface growth in random media, including a summary of recent
developments, can be found. Finally,  a detailed discussion of 
the phenomenology and some theoretical approaches for the study
of MBE growth can be found in the book by
Tu and  Harris (1991)  (see also Tang and Nattermann, 1991; Villain {\it et al.}, 1992; Kessler {\it et al.}, 
1992; Hunt {\it et al.}, 1994; Pal and Landau, 1994; 
Siegert and Plischke, 1994; Das Sarma, 1994; Das Sarma 
{\it et al.}, 1996).

\section{Previous approaches}
\label{two}

In this section we will briefly review the traditional method for 
deriving local growth equations. With very few exceptions
(Meakin {\it et al.}, 1986; Maritan {\it et al.}, 1992), this 
is accomplished within the familiar no--overhang approximation 
(the Monge representation). Denoting by $h(\vec{x},t)$ the 
single--valued interface height function of the lateral coordinates 
$\vec{x}$ and time $t$, one may generally write a growth equation 
in the form
\beq
\frac{\partial h(\vec{x},t)}{\partial t}=G[h(\vec{x},t)] +
\eta(\vec{x},t)
\label{eq21}
\eeq
where $G$ is the deterministic growth term and $\eta$ is the 
noise. This is essentially Newton's law of motion in which the 
inertial term has been neglected in comparison with the 
dissipative force. The neglected term is irrelevant in 
determining the scaling behavior in the asymptotic regime. 
This regime, in Fourier space, is related to the behavior
of correlations functions in the limit of small frequencies
$\omega\to 0$. While the left hand side (LHS) of Eq. \req{eq21}
is proportional to $\omega$, the inertial term is of 
order $\omega^2$ and hence it is irrelevant. 

Consider a $D$ dimensional substrate of size $L^D$ 
(the physical case corresponds to a $L\times L$ substrate with 
$D=2$) and then define the mean height of the growing film and its
roughness by
\begin{mathletters}
\label{eq22}
\beq
\overline{h}(t,L)=\frac{1}{L^D}\int d^Dx h(\vec{x},t),
\eeq
\beq
W(t,L)=\left\langle \frac{1}{L^D}\int d^Dx \left[h(\vec{x},t)-
\overline{h}(t,L)\right]^2\right\rangle^{1/2}
\eeq
\end{mathletters}
where $\avg{\ldots}$ denotes an average over different realizations of the
noise (samples).

Starting from a flat interface (one of the possible initial conditions)
it was conjectured by Family and Vicsek (1985) (see also
Plischke and R\'acz ,1985, Family and Vicsek, 1991)
that a scaling of space by a factor $\ell$ and of time by a factor
$\ell^z$
(dynamic scaling), rescales the roughness $W$ by a factor $\ell^\alpha$, 
as appropriate for a self-affine surface (Mandelbrot, 1986), i.e.,
\beq
W(\ell^z t, \ell L)=\ell^\alpha W(t,L),
\label{scal1}
\eeq
which implies that 
\beq
W(t,L)=L^\alpha f(t/L^z).
\label{scal2}
\eeq
If, for large $t$ and fixed large $L$ ($t/L^z\to\infty$), 
$W$ saturates, then $f(x)\to constant$ as $x\to\infty$.
However for fixed large $L$ and $1\ll t\ll L^z$, one 
expects that correlations of the height fluctuations are set up 
only within a distance $t^{1/z}$ and thus $W$ must be independent of $L$.
This implies that for $x\ll 1$:
\beq
f(x)\sim x^\beta \quad\hbox{with}\quad \beta=\alpha/z.
\label{betadef}
\eeq
Thus dynamic scaling postulates that:
\begin{mathletters}
\label{scal}
\beq
W(t,L)\sim t^\beta  \qquad 1\ll t\ll L^z,
\eeq
\beq
\qquad     \sim L^\alpha \qquad t\gg L^z.
\eeq
\end{mathletters}

The roughness exponents $\alpha$ and the dynamic exponent $z$
characterize the self-affine geometry of the surface
and its dynamics respectively.

The above considerations do not include short time behavior. 
In that limit one may expect that the height fluctuations are
uncorrelated and therefore $h(\vec{x},t)-\overline{h}(t)$
behaves as $\int_0^t \eta(\tau)d\tau$ where $\eta(\tau)$ is a 
white noise with zero average and $\avg{\eta(\tau)\eta(\tau')}
\propto \delta(\tau-\tau')$. This leads to $W(t,L)\sim t^{1/2}$
independent of $L$ and is the random deposition regime.
Note that this dynamic scaling is not related to the standard 
dynamic scaling in critical phenomena. The latter involves 
correlations which are space time separated whereas here they 
concern the behavior at the same time. Furthermore, after the 
interfacial roughness has saturated ($t\gg L^z$), one may expect 
that correlations of fluctuations $\delta h(\vec{x},t) = h(\vec{x},t) - 
\overline{h} (t,L)$ such as:
\beq
C_L(\vec{x},t;\vec{x}',t')=\left\langle
\left[\delta h(\vec{x},t)-\delta h(\vec{x}',t')\right]^2\right\rangle
\label{cl}
\eeq
show the more traditional dynamic scaling:
\beq
C_L(\vec{x},t+\tau;\vec{x}',t)=|\vec{x}-\vec{x}'|^{2\alpha}
g\left(\frac{\tau}{|\vec{x}-\vec{x}'|^{z'}},
\frac{|\vec{x}-\vec{x}'|}{L}\right)
\label{scalcl}
\eeq
with $z'$ not necessarily equal to $z$. Note that the exponent 
$\alpha$ must be the same as before since:
\beq
\frac{1}{L^{2D}}\int d^Dx d^Dx'C_L(\vec{x},t;\vec{x}',t)=
W^2(t,L)\sim L^{2\alpha} \qquad (t\gg L^z).
\label{defalpha}
\eeq
The key idea behind the traditional
derivation of $G$ and $\eta$ in Eq. 
\req{eq21} is 
the identification of the symmetries and conservation laws of the 
system (Kardar, 1994). 
The deterministic term $G$ is expanded in powers and 
combinations of the $h$ field and the relevant (in the 
renormalization group sense) lowest order terms that are 
consistent with the symmetries and conservation laws are retained.
The lowest order contribution is a constant $G_0$ which can
always be set to zero by the transformation $h\to h+G_0t$. This 
amounts to the use of a  reference frame that co-moves with the
interface. 
Symmetries that need to be considered include invariance under 
translation in time and space, and along the growth direction, rotational 
and inversion symmetry about the growth direction. 
For example, translational invariance in the growth direction, 
which holds in most practical cases, rules out terms proportional 
to powers of $h$. An explicit dependence of $G$ on time 
or position $\vec{x}$ contradicts time or space translational invariance, 
which usually holds, and can therefore be neglected.
In addition it is helpful to classify terms 
which might occur in the expansion in two categories: those which 
conserve the number of particles and those which do not.
To derive the equation of growth for a conservative dynamics, 
the key observation (Villain, 1991; Kardar, 1994) is that the 
deterministic part must have the form of a continuity equation:
\beq
\frac{\partial h(\vec{x},t)}{\partial t}=
-\vec{\nabla}\cdot\vec{j}(\vec{x},t),
\label{contin}
\eeq
where the macroscopic current $\vec{j}
(\vec{x},t)$ describes the flux of atoms on the surface.
The current $\vec{j}(\vec{x},t)$ arises in general from
differences in the local chemical potential $\mu(\vec{x},t)$.
The relation between $\vec{j}(\vec{x},t)$ and $\mu(\vec{x},t)$:
\beq
\vec{j}(\vec{x},t)=-\vec{\nabla}\mu(\vec{x},t)
\label{jmu}
\eeq
describes the fact that atoms drift to regions of 
minimum chemical potential.
The simplest source of chemical potential, which is 
gravitational energy $\mu\propto h$, led 
Edwards and Wilkinson (1982) to formulate the equation
\beq
\frac{\partial h(\vec{x},t)}{\partial t}=\nu
\nabla^2h(\vec{x},t) + \eta(\vec{x},t).
\label{ew}
\eeq
The noise here arises due to the stochastic nature of the incoming 
flux of atoms and, as such, it is non--conservative.
The noise can also result from the probabilistic 
nature of the diffusion process on the surface or from thermal fluctuations.
In the latter cases, the number of atoms on the surface 
remains the same and the noise will be called conservative. 
The noise correlation functions are different in the two cases and are given by:
\beqar
\avg{\eta(\vec{x},t)\eta(\vec{x}',t')}&=& 2 D_0\delta^D(\vec{x}-
\vec{x}')\delta(t-t')\qquad\hbox{non--conservative}
\label{nonconsnoise}\\
\avg{\eta(\vec{x},t)\eta(\vec{x}',t')}&=& (-D_1\nabla^2+D_2\nabla^4)
\delta^D(\vec{x}-\vec{x}')\delta(t-t')\qquad\hbox{conservative}
\label{consnoise}
\eeqar
We shall not consider the case where the statistics of $\eta$ is not
gaussian (see e.g. Zhang 1990, and Krug 1991, Horv\'ath {\it et al}, 
1991, Lam and Sander 1993) or that in which it is long range correlated
in space or in time (see Medina {\it et al.} 1989, 
Peng {\it et al.} 1991, Lam {\it et al.} 1992).

Usually,  the effect of gravity on deposition processes 
is totally negligible with respect to other effects. 
However,  the Edwards--Wilkinson equation also applies  
to other situations and mainly to the case where 
atoms are allowed to evaporate from the surface. 
This effect, which does not conserve the number of particles, 
is nevertheless described by a term $\nu \nabla^2h$ in the
growth equation. We repeat here the derivation, given by
Villain (1991). In general, if 
the deterministic growth term $G$ describes  dynamics
which minimizes a potential $V[h]$ (which is a functional
of the interface configuration $h(\vec{x},t)$) it will have the 
form
\beq
G=-\nu \frac{\delta V(h)}{\delta h(\vec{x},t)}
\label{potential}
\eeq
where $\delta/\delta h(\vec{x},t)$ denotes a functional 
derivative. The effect of evaporation is to
minimize the surface area. Since the excess surface area due to
roughness is given, in a small gradient expansion, by $V[h]=\int d^dx 
\left(\vec{\nabla} h\right)^2$ (Bruinsma and Aeppli, 1984), 
Eq. \req{potential} directly gives the first term 
on the right hand side of Eq. \req{ew}.
It is interesting to note already at this stage how
a relevant feature of the process, i.e.,  whether the number 
of particles is conserved or not, is lost in the small
gradient expansion of the growth equation. 
A derivation of the growth equation from the R--invariance principle 
avoids such ambiguities.

Let us return to conservative equations and look for
the simplest equation one gets by neglecting gravity.
>From the above discussion,  it is natural to expect that
the next term in the expansion of the chemical potential 
will be a term proportional to the local curvature 
$\mu\propto \nabla^2h$. This term, via Eqs. \req{contin}, \req{jmu}, 
will favour again a flux of particles away from the local 
maxima towards the local minima. Hereafter, the resulting relaxation 
mechanism will be called surface diffusion (Villain 1991;
Siegert and Plischke, 1992, 1993; Das Sarma {\it et al.}, 1996).
Since this dynamics constrains the atoms to stay on the surface,
this relaxation will be much 
slower than the one provided by evaporation. The resulting
linear equation is
\beq
\frac{\partial h(\vec{x},t)}{\partial t}=-\kappa
\nabla^4 h(\vec{x},t) + \eta(\vec{x},t)
\label{linmbe}
\eeq
and the same considerations discussed previously
apply to the noise term here.
The two equations can be combined into one which accounts
both for gravity and for surface diffusion. In experimental
situations of MBE, however, the typical coefficients $\nu$ and
$\kappa$ are such that the effects of gravity are relevant
only on length scales much bigger than the typical sample size.
We note that both terms discussed so far satisfy all the
symmetry properties mentioned previously. Note that while rotational
invariance in the substrate plane is satisfied by these terms,
the full space rotational invariance is lost. 
This is a consequence of the small gradient expansion in $h$
rather than a physical characteristic of the processes. 
For example, atoms diffuse on the interface in the same 
manner irrespective of the direction we assign to the $\hat{z}$
axis. Again, we will recover the full space rotational invariance,
and distinguish terms which satisfy this symmetry and those
that do not, once the R--invariant form of the equation is derived.

The lowest order non--linear term which one could include 
in $G$ leads to the Kardar-Parisi-Zhang (KPZ) equation (Kardar {\it et al.},
1986)
\beq
\frac{\partial h(\vec{x},t)}{\partial t}=\nu\nabla^2h(\vec{x},t) 
+\frac{\lambda}{2} (\vec{\nabla}h)^2 + \eta(\vec{x},t).
\label{kpz}
\eeq
Here the mechanism of relaxation of surface fluctuations is 
the same as that of the EW equation \req{ew}. 
The origin of the non--linear term lies in the driving force of the 
deposition process that is perpendicular to the interface (or 
lateral growth in the discrete ballistic deposition model -- for a review
of ballistic deposition, see Meakin,  1993.  See also Ko and Seno, 1994
for an off-lattice simulation).
While the simplest growth term, a constant $G_0$, can
be eliminated by choosing a co-moving frame on the interface
(see above), the non--linear term of the KPZ equation cannot. 
The resulting dynamical process is intrinsically 
irreversible.   Interesting generalizations of the KPZ equation to
a multi-component model have been 
studied by Doherty {\it et al.} (1994).

Among the notable properties of the KPZ equation are its
relations with a large number of other problems, including 
the statistics of directed polymers in random media 
(Kardar and Zhang, 1987, Fisher and Huse, 1991,
Kim {\it et al.}, 1991, Halpin-Healy and Zhang, 1995)
and the Burgers equation for fluid dynamics 
(Forster {\it et al.}, 1977). The Galilean
invariance in the latter problem has profound consequences 
on the properties of the KPZ equation, and leads to 
the exponent identity $z+\alpha=2$
(Forster {\it et al.}, 1977; Meakin {\it et al.}, 1986). This invariance 
translates the problem of interface growth into an 
invariance of the equation for an infinitesimal tilt 
of the substrate plane. This invariance can be seen as a 
remnant of the full $D+1$ rotational invariance of a
growth process occurring in the direction normal to the 
interface. This feature can be fully appreciated
in the R--invariant form (Maritan {\it et al.}, 1992) 
of the equation which describes 
an isotropic growth process driven by a pressure
or occurring from the condensation of a vapor.
In the gradient expansion leading to Eq. \req{kpz},
rotational invariance is retained only for 
infinitesimal transformations. 
However, as we shall see, there are other mechanisms,
which are not rotationally invariant, which lead to a 
growth equation of the form \req{kpz} in the small gradient 
expansion. 

Note that in the KPZ equation the $h\to -h$ symmetry is broken:
there is a definite growth direction. Also the KPZ equation 
describes processes in which the number of particles is 
not conserved. The $(\vec{\nabla}h)^2$ term cannot appear in
a situation, such as  Molecular Beam Epitaxy (MBE), 
where conservation of the number 
of particles on the interface is expected to hold.

One may invoke dynamic scaling to show that, in general, an 
equation of the type: 
\[\frac{\partial h}{\partial t}=(-1)^{(n+1)} K_n (\nabla^2)^n h +\eta
\quad\hbox{with}\quad\avg{\eta(\vec{x},t)\eta(\vec{x}',t')}=
2D\Gamma_m\delta^D(\vec{x}-\vec{x}')\delta(t-t'),\]
where $m$ takes on the two values $0$ or $1$ with 
$\Gamma_0=1$ and $\Gamma_1=-\nabla^2$, has the exponents
\[z=2n\quad\hbox{and}\quad \alpha=\frac{2(n-m)-D}{2}.\]
Thus for the Edwards--Wilkinson model Eq. \req{ew} and for
Eq. \req{linmbe}, one obtains ($m=0$ in both cases)
\[z=2,\quad \alpha=\frac{2-D}{2}\qquad\hbox{and}\qquad
z=4,\quad \alpha=\frac{4-D}{2}\]
respectively.
The situation with $\alpha <0$ corresponds to $W(t\to\infty,L)
\sim constant$ ($\alpha=0$ leads to logarithmic corrections).
Since the continuum equation 
was derived in a power series of gradients of $h$ with only the linear
term being retained, the result $\alpha\ge 1$ is a matter of concern,
since local slopes on a distance of order $\ell$ scale as $|\nabla h|\sim 
\ell^{\alpha-1}$. This suggests that neglected non--linear terms 
need to be considered as well (for a discussion on
self-similar ($\alpha=1$) and self-affine surfaces
see Mandelbrot, 1986 and Meakin {\it et al}, 1986). 
The requirements that the basic symmetries 
be obeyed and that in the surface diffusion dominated regime the 
continuity equation \req{contin} be satisfied lead to:
\beq
\frac{\partial h(\vec{x},t)}{\partial t}=
\nu\nabla^2 h(\vec{x},t) 
-\kappa\nabla^4 h(\vec{x},t) 
+\lambda_1 \nabla^2(\vec{\nabla}h)^2 
+\lambda_2 \vec{\nabla}[\vec{\nabla}h(\vec{\nabla}h)^2] 
+ \eta(\vec{x},t),
\label{nonlinmbe}
\eeq
in which all the terms up to fourth order in the gradient expansion 
of $h$ have been retained. 
Physically $\nu$ is set to zero since it arises from a 
gravity--like chemical potential which, as mentioned above and as we 
will discuss below, is negligible. 
The requirement that the surface current $\vec{j}$ results
from a chemical potential would rule out the $\lambda_2$ term,
even though this term is more relevant, from a renormalization
group point of view, than the $\lambda_1$ term.
It will be clear, from the R--invariant form of the equation, that the
$\lambda_1$ term arises as the second one in an expansion
of a gravity or surface tension term, the first one being 
the $\nu$ term. It is interesting to note that recently (Das Sarma 
and Kotlyar, 1994; Kim and Das Sarma, 1995)
it has been proved that, even if the coefficient of the Laplacian is
zero in the starting equation, a $\nu>0$  is generated
under renormalization  by the $\lambda_2$
term, leading ultimately to the EW behavior for surface 
fluctuations. 

Recently a  growth equation has been derived by Vvedensky {\it et al.}, (1993)
starting from a microscopic solid--on--solid model with deposition,
desorption and diffusion through a master equation for the surface
dynamics. In the continuum limit, the growth equation is obtained 
-- desorption leads to non--zero $\nu$ and
$\lambda$ values while surface diffusion produces the $\kappa$ term.

Similar considerations have been used to derive growth equations
for interfaces in a disordered medium (Koplik and Levine, 1988;
Jensen and Procaccia, 1991; Parisi, 1992; Nattermann {\it et al.}, 1992; 
Jiang and Hentschel, 1992; Sneppen, 1992; Nolle {\it et al.}, 1993; 
Olami {\it et al.}, 1994; Galluccio and Zhang, 1995)
such as domain walls in random field Ising models (Nattermann, 1985; 
Fisher, 1986; Ji and Robbins, 1992; Narayan and Fisher, 1993) and
the invasion of one fluid into another within a porous medium 
(Nittmann {\it et al.}, 1985; Stokes {\it et al.}, 1988;
Buldyrev {\it et al.}, 1992; He {\it et al.}, 1992; Tang and Leschhorn, 1992; 
1993; Leschhorn, 1993; Delker {\it et al.}, 1996). 
The disorder serves 
to pin parts of the interface. This feature is captured by replacing the 
thermal noise with a quenched random noise generated by the disorder 
$\eta(\vec{x},h)$ governed by correlations of the form
\beq
\avg{\eta(\vec{x},h)\eta(\vec{x}',h')}= 2 D_0\delta^D(\vec{x}-
\vec{x}')\Delta(h-h')
\label{disordersnoise}
\eeq
where $\avg{\ldots}$ represents an average over different 
realization of the randomness and the function $\Delta$ 
characterizes the nature of the quenched disorder.
It is important to note that disorder will break the 
translational invariance in the $h$ direction. Hence a 
constant $G_0$ term can no longer be eliminated by choosing
the co-moving frame, since under the transformation 
$h(\vec{x},t)\to h(\vec{x},t)+G_0 t$, it will reappear in the argument of the 
noise $\eta(\vec{x},h)\to \eta(\vec{x},h+G_0 t)$. 
Here, $G_0$ plays the role of the driving force: for 
small $G_0$ the interface will find sooner or later 
a surface where the values of the pinning forces are 
strong enough to inhibit further growth. 
In contrast, if $G_0$ is very large, 
$h$ in the argument of the noise in the co-moving frame 
can be neglected with respect to $G_0 t$, and one expects 
to recover the behavior of the corresponding dynamics
without disorder, i.e., $\Delta(h-h') \propto \delta(t-t')$. 
In practice the interface will move
so fast that it will sample so many values of the disorder 
in a small time interval, that the overall effect is
that of a time dependent noise with correlations of the
form \req{nonconsnoise}. We shall not deal in the following 
with surface growth in disordered media, a steadily growing
subject which is  reviewed in the book by Barab\'asi
and Stanley (1995).

\section{Equations for a Growing Surface and 
Reparametrization Invariance}
\label{three}

The most general Langevin equation for the evolution of a surface
in a $D+1$ dimensional space has the form

\begin{equation}
\partial_t\vec{r}(\U{s},t)=\hat{n}(\U{s},t) {\cal G}[\vec{r}(\U{s},t)]
+\vec{F}(\U{s},t),
\label{eq1}
\end{equation}        
where  the   $D+1$     dimensional   vector
$\vec{r}(\U{s},t)   \equiv\{r_\alpha(\U{s},t)\}_{\alpha=1}^{D+1}$
runs over the surface as  $\U{s}\equiv\{s^i\}_{i=1}^D$, varies in
a  parameter space. (See Appendix \ref{appA} for a brief summary of 
the elements of differential geometry). 
In Eq.  \req{eq1}  $\hat{n}$  stands for the
versor  normal  to the  surface  at  $\vec{r}$  while  ${\cal G}$
contains  a  deterministic  growth  mechanism that  causes growth
along the normal $\hat{n}$ to  the surface and is a functional of
$\vec{r}$   itself.  $\vec{F}$ is a  random  force  acting on the
surface. Eq. \req{eq1} derives from Newton laws in the limit of
a massless surface when the inertial term $\partial_t^2\vec{r}$ can be
ignored with respect  to the dissipative  force. Note that
the time derivative  of $\vec{r}$ has to be  parallel to the normal
to the surface. This is  because $t=s^0$ can  be
regarded as the $D+1^{\rm th}$  coordinate, and  $(\U{s},s^0)$ is a
curvilinear   coordinate  system. If the
growing surface invades the $D+1$ dimensional space, this
parametrization is legitimate since the metric tensor is positive
definite. However  $s^0$ is  the absolute
time,  and   changes of    parametrization  cannot   involve this
variable. This is satisfied  only if the elements $g_{0,i} \equiv
\partial_t\vec{r} \cdot {\partial_i}\vec{r}$ vanish, which 
implies  that $\partial_t\vec{r}\perp{\partial_i}\vec{r}$
for all $i=1,\ldots,D$, and therefore $\partial_t\vec{r}\parallel 
\hat{n}$. 

Independent of
the physical mechanisms entering the various terms of Eq.
\req{eq1}, which specify the form of ${\cal G}$ and the
properties   of  $\vec{F}$,  this   equation has  to  satisfy the
fundamental     requirement  of    reparametrization   invariance
(R--invariance).   This  requires that  only  quantities that are
independent of the  choice of the  parametrization $\U{s}$, such
as those referring to the local geometry of the surface, like the
curvature,  can appear in  the  equation. As with any  other symmetry,
reparametrization  invariance  poses  constraints on the possible
forms that  Eq. \req{eq1}  can take. 

The basic elements of  differential geometry used in the 
derivations and some calculational details are presented in
the appendix.

It will be convenient to express the  deterministic part 
of Eq.  \req{eq1}, as  well as the noise, 
as a  sum of  different  terms 
\[{\cal G}={\cal G}_a+{\cal G}_b+\ldots~~~\hbox{and}~~
\vec{F}=\vec{F}_a+\vec{F}_b+\ldots.\]     
The   derivation of the
equation can  then be split into those of the individual
terms that are expected to appear in a given physical system.

It is often convenient and sufficient to
describe the interface in the Monge form, i.e., with
$\vec{r}=(\U{x},h(\U{x}))$. 
Henceforth, at variance with section \ref{two}, we will use the 
notation $\U{x}$ instead of $\vec x$ to stress that the Monge 
form is one of the many parametrizations one can use.
$h(\U{x})$ is the coordinate in the
direction normal to the substrate and since it must be a single
valued function of $\U{x}$ the interface cannot have overhangs.
In this form \req{eq1} becomes

\begin{equation}
\partial_t h(\U{x},t)=\sqrt{g}\left({\cal G}+\eta\right)
\label{mongeq1}
\end{equation}
where $g=1+(\U{\nabla h})^2$ is the determinant of the metric 
tensor  and the relation
$\hat n= (-\U{\nabla} h,1)/\sqrt{g} $ has been used
(see Appendix \ref{appA}). 
${\cal G}$ and $\eta=\vec{F}\cdot\hat{n}$, as before, are the amplitude
of the deterministic force and the noise 
in the normal direction respectively
(note that $\partial_t h(\U{x},t)/\!\sqrt{g}$ is the normal
velocity $ \hat n \cdot \partial_t \vec{r} $ -- this is easy 
to prove using Eq. \req{mongelast}).
The discussion of the noise term is presented in Section 
\ref{three.B}.

\subsection{Deterministic evolution }
\label{three.A}

As stressed before, reparametrization invariance  requires 
that ${\cal G}$ depends
only on intrinsic geometric  properties of the interface such as the
mean  curvature  $H$ or, when ${\cal G}$ is not rotationally invariant, 
on  scalar  products of the
normal $\hat{n}$  with some fixed vector  $\vec{v}$.

The physical meaning of ${\cal G}$ is particularly evident in the
case in which it can be derived from a potential:

\begin{equation}
\left.\partial_t\vec{r}(\U{s},t)\right|_{\det}=-\frac{1}{\sqrt{g(\U{s})}}
\frac{\delta {\cal H} [\vec{r}(\U{s})]}{\delta\vec{r}(\U{s})},
\label{eq2}
\end{equation} where ${\cal H}$  is an R--invariant functional of
$\vec{r}$. 
The $ 1/ \sqrt{g} $ term in Eq. \req{eq2} appears since 
$\frac {1}{\sqrt{g(\U{s})}} \frac {\delta f (\U{s'})}{\delta f (\U{s})}=
\frac{ \delta (\U{s}-\U{s'})}{\sqrt{g(\U{s})}}$
is the R--invariant Dirac's delta function (see Appendix A).
In this  case, the dynamics 
tends to minimize the potential energy ${\cal H}$ of the surface.
Moreover, if the  random force is properly  chosen
(this will be discussed in section \ref{three.B}), i.e., if it is
not  conservative,  the system  approaches  a steady  state whose
distribution   of   $\vec{r}$  is  given by    $\exp\{-\beta{\cal
H}[\vec{r}]\}$  where $\beta$  is related to  the correlations of
the noise (Hohenberg and Halperin, 1977).

The R--invariance   of  ${\cal H}$   guarantees  that the  functional
derivative in Eq.\req{eq2} is a vector parallel to the normal as required
by Eq. \req{eq1}, and
the $1/\sqrt{g}$ factor  guarantees R--invariance of the functional
derivative.   Indeed  R--invariance  of  ${\cal H}$  implies that
${\cal   H}[\vec{r}(\U{s}')] ={\cal   H}[\vec{r}(\U{s})]$ for any
reparametrization      $\U{s}'(\U{s})  $.  
For an   infinitesimal transformation 
$\U{s}'=\U{s}+\U{\epsilon}(\U{s})$,
\[\vec{r'}(\U{s}) \equiv \vec{r}(\U{s}')=\vec{r}(\U{s})+
\U{\epsilon}(\U{s})\cdot\U{\partial}\vec{r}(\U{s})+O(\epsilon^2)\]
so that
\[{\cal H}[\vec{r}(\U{s}')]={\cal H}[\vec{r}(\U{s})]
+\int d^D s \epsilon^i(\U{s}) {\partial_i}\vec{r}(\U{s})\cdot
\frac{\delta {\cal H}}{\delta\vec{r}(\U{s})}.\] 
Since  ${\partial_i}\vec{r}$ is a vector in
the tangent plane of the surface, the second  term in this equation 
vanishes for any function $\U{\epsilon}$ only if the functional derivative is
parallel to the normal $\hat{n}$ .

In the Monge form,
if ${\cal G}$ derives from a potential ${\cal H}$ we find
(see appendix \ref{appB}):

\begin{equation}
{\cal G}=-\hat{n}\cdot\frac{1}{\sqrt{g}}\frac{\delta 
{\cal H}}{\delta\vec{r}}
=-\frac{\delta {\cal H}}{\delta h}.
\label{dhmonge}
\end{equation}

We now proceed to consider the  simplest possible  terms that can
appear  in Eq.   \req{eq1}. We  first  give the   expression in a
general parametrization and  then discuss the Monge form and the
expansion in small gradients of $h(\U{x},t)$.

\subsubsection{Surface tension}
\label{three.A.1}

The simplest  physically  motivated term in the  Hamiltonian of a
surface  is    proportional to  the  total  area  ${\cal  A}=\int
d^Ds\sqrt{g}$  and  produces a force  that tends to  minimize the
surface  area. This term  is usually  referred to  as arising from
surface tension. The functional derivative yields (see Appendix 
\ref{appC})

\begin{equation}
-\frac{1}{\sqrt{g}}\frac{\delta}{\delta\vec{r}(\U{s})}\int
d^Ds\sqrt{g}=\frac{1}{\sqrt{g}}{\partial_i}\left(\sqrt{g} g^{i,
j}\partial_j\right)\vec{r}(\U{s}) =\bl \vec{r}(\U{s})
\label{srf}
\end{equation}
where $\bl$ is the Beltrami--Laplace operator defined in Eq. \req{bldef}.
For ${\cal H}_s=\nu_s {\cal A}$, using \req{Hdef2}, one gets: 

\begin{equation}
{\cal G}_s=\nu_s\hat{n}\cdot\bl\vec{r}= \nu_s H,
\label{surf}
\end{equation}
where $H$ is the mean curvature.

Using Eqs. \req{dhmonge}, \req{surf} and 
$ {\cal H}_s = {\nu}_s \int d^D x \sqrt{ 1 + (\unabla h)^2} $,
one finds, in a small gradient expansion in the Monge parametrization (Eq. 
\req{mongeq1}):

\begin{equation}
\sqrt{g}{\cal G}_s=\nu_s\sqrt{g}\U{\nabla}\cdot
\frac{\U{\nabla} h}{\sqrt{g}}= 
\nu_s\left[\nabla^2 h(\U{x},t) +\frac{1}{2} \nabla^2 h(\unabla h)^2
-\frac{1}{2} \U{\nabla}\cdot[\U{\nabla}h(\U{\nabla}h)^2] 
+ \ldots\right].
\label{srfmg}
\end{equation}

The main physical mechanism which produces a term like this is 
evaporation. Following Villain (1991) we observe
that the evaporation rate will be proportional to
the difference between the chemical potential of the solid
$\mu_s$ and that of the vapor $\mu_v$. On the 
surface, the former will depend on the local geometry and hence:
\[\partial_t\vec{r}|_{\rm evap.}
=\hat{n} B(\mu_s(\vec{r})-\mu_v).\]
We can then expand $\mu_s$ in powers of the local curvature.
Along with the zeroth order term, corresponding to the chemical potential for 
a flat surface, there will appear a term proportional to
the curvature $H$ as in the r.h.s of Eq. \req{surf}.

\subsubsection{Pressure}
\label{three.A.2}

The second simplest geometrical property on which ${\cal H}$
may depend is the volume  enclosed by the surface, which may be written
as ${\cal V} = {1\over{D+1}} \int d^Ds \sqrt{g(\U{s})}\vec {r}(\U{s})
\cdot \hat n(\U{s})$.
A linear dependence of ${\cal H}$ on the volume physically represents a
pressure term. If ${\cal H}_v=-\lambda {\cal V}$, the
pressure $\lambda>0$ encourages an increase in the volume, while if
$\lambda<0$ the force in Eq. \req{eq2} acts to deflate the surface.
The infinitesimal volume variation on the surface element
$d\sigma=d^Ds\sqrt{g}$ is given by $d\sigma
\hat{n}\cdot\delta\vec{r}$, so the functional derivative of ${\cal
H}_v$ in Eq. \req{eq2} gives:

\begin{equation}
{\cal G}_v=-\hat{n}\cdot\frac{1}{\sqrt{g}}
\frac{\delta {\cal H}_v}{\delta\vec{r}}=\lambda,
\label{vol}
\end{equation}
that in the Monge form becomes

\begin{equation}
\sqrt{g}{\cal
G}_v=\lambda\sqrt{g}=\lambda\left[1+\frac{1}{2}(\U{\nabla}h)^2+
\ldots\right].
\label{volmg}
\end{equation} 

Equation \req{eq1} with ${\cal G}={\cal G}_s+{\cal G}_v$ is 
one of the R--invariant forms (Maritan {\it et al.}, 1992) of the KPZ equation: 
\beq 
\left.\partial_t\vec{r}(\U{s},t)\right|_{\det}=\hat{n} ({\nu}_s H + \lambda).
\label{RIKPZ}
\eeq
which in the Monge representation reads:
\beq 
\left.\partial_t h(\U{x},t)\right|_{\det}=
{\nu}_s\sqrt{g}\U{\nabla}\cdot
\frac{\U{\nabla} h}{\sqrt{g}} + \lambda\sqrt{g}.
\label{MKPZ}
\eeq
Indeed, to lowest
order in the gradient expansion, Eq. \req{kpz} is recovered (apart from 
the  constant term  $\lambda$ which can be absorbed by redefining 
$h\to  h+\lambda  t$).
The  complete   R--invariant KPZ  equation  derives from the
Hamiltonian:
\beq
{\cal  H}_{{}_{KPZ}}=\int d^Dx(\nu_s\sqrt{g}-\lambda h),
\label{hkpz}
\eeq
which, however, is unbounded as $h\to\infty$. 
Thus, even with a suitable noise term, Eq. \req{MKPZ} does not 
have $\exp (-\beta  {\cal H}_{{}_{KPZ}})$
as the equilibrium distribution of $h$.
This is  not surprising  and is  related to the
presence of a  pressure that makes the  system grow forever. In a
sense, the  interface growth is  intrinsically  irreversible. Note
that $\lambda$ couples only to the $k=0$ mode of $h$ in a Fourier
expansion. This would suggest that  steady state correlation
functions, such as  $\avg{[h(\U{x})-h(\U{y})]^2}$, are independent of
$\lambda$. A  derivation of the KPZ  equation from the functional
derivative of a free energy with a volume and a surface term, was
also obtained  by Grossmann {\it et al.} (1991) in a 
more complex way.  An alternative derivation was also given in Keblinski
{\it et al.}, (1996).

\subsubsection{Curvature energy}
\label{three.A.4}

The potential ${\cal H}$ may also depend on
the curvature $H$ of the interface. In general, this dependence
can be expressed in a power series expansion  (the zeroth order term 
has already been considered in Section \ref{three.A.1})

\begin{equation}
{\cal H}_c=\int d^Ds\sqrt{g}(\kappa_1 H+\kappa_2
H^2+\ldots)={\cal H}_{c,1}+{\cal H}_{c,2}+\ldots.
\label{hc}
\end{equation}
The physics behind the first term reflects the difference in the
mechanical properties of the media divided by the
interface. Indeed for $\kappa_1>0$,  large negative curvatures are
encouraged while positive  ones are depressed. The functional
derivative is carried out in the appendix \ref{appD} with
the result
\begin{equation}
{\cal G}_{c,1}=-\frac{1}{\sqrt{g}}\hat{n}\cdot
\frac{\delta {\cal H}_{c,1}}{\delta\vec{r}}=\kappa_1 
\left(H^2-\sum_{i=1}^D \lambda_i^2\right),
\label{curv1}
\end{equation}   
where  $\lambda_i$  are the   eigenvalues of the
matrix of  the  coefficients of the  second  fundamental form and
express the principal curvatures of the surface. 
Since $H=\lambda_1$ in  $D=1$, ${\cal G}_{c,1}$
vanishes. This is a consequence of the
Gauss--Bonnet   theorem  that  states  that the   integral of the
Gaussian  curvature $K$  on a closed  surface is a constant.
Since $H=K$  in $D=1$, the  variation of ${\cal H}_{c,1}$
is zero.

In terms of $h(\U{x})$, Eq. \req{curv1} takes the form:

\begin{equation}
\sqrt{g}{\cal G}_{c,1} 
\simeq
\kappa_1 \left[(\nabla^2 h)^2-
\sum_{i,j=1}^D (\partial_i \partial_j h)^2+\ldots\right].
\label{curv1mg}
\end{equation}  
In the  small  gradient expansion  no linear term in $h$
arises. The curvature term, as expected, breaks the symmetry 
$h\to -h$ in the growth equation.

Similarly, higher powers of $H$ are easily worked out ( see Appendix
\ref{appD}). For the $p^{\rm th}$ term in Eq. \req{hc} we
find:

\begin{equation}
{\cal G}_{c,p}=-\frac{1}{\sqrt{g}}\hat{n}\cdot\frac{\delta 
{\cal H}_{c,p}}{\delta\vec{r}}= \kappa_p
\left(H^{p+1}- p\bl H^{p-1}- pH^{p-1}\sum_{i=1}^D \lambda_i^2\right).
\label{curvp}
\end{equation} This equation  generalizes Eq. \req{surf}, that is
the $p=0$ case, and Eq. \req{curv1} ($p=1$). The effect of these terms is
more  transparent in the  Monge  parametrization.  Of particular
interest  is the  $\kappa_2$ term in  the  expansion of the
curvature     potential:
\beq
\sqrt g {\cal G}_{c,2}
=-2\kappa_2(\nabla^2)^2  h + \ldots
\label{curv2mg}
\eeq

\subsubsection{Rotational Invariance}
\label{three.A.3a}

All the $\cal G$'s we have considered so far are invariant under 
rotations in the $D+1$ space.  This implies (unless the noise term 
breaks this invariance) that the evolution equation in its form Eq. 
\req{mongeq1} is invariant under the following (infinitesimal) symmetry 
transformations:

\[\underline{x}' = \underline{x} - \underline \epsilon h (\underline{x},t),\]

and

\[h'(\underline{x}',t) = h(\underline{x},t) + \underline{\epsilon} \cdot 
\underline{x}.\] 

\noindent
Indeed, it is straightforward to verify that the normal velocities 
$\frac{\partial _{t}h(\underline{x},t)}{\sqrt{g(\underline{x},t)}}$ and 
$\frac{\partial _{t}h'(\underline{x}',t)}{\sqrt{g'(\underline{x}'t)}}$ are
equal.

\noindent
In the small gradient expansion of Eq. \req{MKPZ}, Eq. \req{kpz} is 
obtained after 
redefining $h\rightarrow h + \lambda t$.   As a consequence, the above symmetry 
transformations become

\[\underline{x}' = \underline{x} - \underline{\epsilon} \lambda t,\]

and

\[h'(\underline{x}',t)=h(\underline{x},t) + \underline{\epsilon} \cdot 
\underline{x}.\]

\noindent
where $-\underline{\epsilon} h(\underline{x},t)$ has been omitted in  the first
equation,  since it contributes only  to the higher order powers neglected in 
Eq. \req{MKPZ}.  These  symmetry transformations are exact for the KPZ 
equation and  can also be   deduced from the Galilean invariance of the 
related Burgers' equation   (Forster {\it et al.}, 1977; Halpin-Healy and Zhang,
1995).

\subsubsection{External potentials}
\label{three.A.3}

In the presence of an external potential, ${\cal H}$ depends
explicitly on $\vec{r}$.  The simplest example is
that of a gravitational field.  The variation $\delta {\cal H}_g$
of the gravitational energy,  when 
$\vec{r}\to\vec{r}+\delta\vec{r}$,      is    proportional to the
variation of the mass $\rho
d^Ds\sqrt{g}\hat{n}\cdot\delta\vec{r}$  ($\rho$ is the 
mass density) multiplied by  the  acceleration
$a_g$  of  the gravitational  field and  the   ``height''
$\hat{z}\cdot\vec{r}$,  where  $\hat{z}$ is the  direction of the
gravitational field. Then

\begin{equation}
{\delta {\cal H}}_g=\nu_g\int
d^Ds\sqrt{g}(\vec{r}\cdot\hat{z})\hat{n}\cdot\delta\vec{r}.
\label{Hgrav}
\end{equation} 
where  $\nu_g=\rho a_g$. It is easy to see that
such a term   breaks   translational invariance, i.e., 
the growth equation   changes if $\vec{r}(\U{s})\rightarrow
\vec{r}(\U{s})+\vec{r}_0$. This generally applies to any potential 
$V(\vec{r})$ for which ${\cal H}=\int d^Ds\sqrt{g}V(\vec{r})$, 
unless $V(\vec{r})=\nu_s$ which yields a surface tension term (see 
\ref{three.A.1}), or $V(\vec{r})=-\lambda \hat n \cdot \vec r /(D+1)$, 
corresponding to the case of a pressure potential
dealt with in \ref{three.A.2}. 
Since translational invariance is expected to  hold, these  terms 
are not considered. In the Monge
parametrization, the  gravitational  energy, proportional to
$\int  d^Dx  h^2(\U{x})$, leads to a linear term in $h$ in the
equation of motion.

\subsubsection{Orientational energy}
\label{three.A.5}

The simplest, translational invariant term which breaks rotational 
invariance, results from considering a potential which depends 
on the local orientation of the surface:
\begin{equation}
{\cal H}_z=-\int d^Ds\sqrt{g}\chi(n_z),
\label{hnz}
\end{equation} 
where $n_z=\hat{n}\cdot\hat{z}$, $\hat{z}$ is some
fixed direction in the $D+1$ dimensional space and $\chi(x)$ is a
generic   function. Such a  term  would  result, for  example, by
imposing a  constraint on the slopes of  the surface with respect
to a   reference  substrate  plane.  This  is  often  realized in
restricted  solid on  solid (RSOS)  models for  interface growth
(Kim and Kosterlitz, 1989).
These are lattice models in  which the height of the surface from
the reference plane at two neighboring sites can differ only by a
given number of  units, for example, $\pm 1$. The effect of such a
constraint propagates at distances larger than the lattice spacing
by discouraging configurations with large slopes (in view of their
low ``entropic'' weight). Hence, at a  coarse grained level, the
effect of the constraint can be expressed by  a term derived 
from Eq. \req{hnz} in a continuum description.  
The functional derivative 
of Eq. \req{hnz} is performed in detail in Appendix \ref{appE}
with the result:
\begin{eqnarray}
{\cal G}_{z}
&=&
-\frac{1}{\sqrt{g}}\hat{n}\cdot\frac{\delta 
{\cal H}_z}{\delta\vec{r}} \nonumber \\
&=& 
H \left[n_z\frac{d\chi(n_z) }{dn_z} - \chi(n_z)\right]
+ \frac{d^2 \chi(n_z) }{d{n_z}^2} \partial^i (\hat z \cdot \vec r)
\partial_i n_z.
\label{Gz}
\end{eqnarray}

The first term in Eq. \req{Gz} is the one we would obtain 
by assuming that 
$d^Ds \sqrt g \hat n \cdot \hat z $,
the projection of an infinitesimal area onto the substrate plane,
does not change under the 
transformation $\vec r \to \vec r + \delta \vec r$.
In the Monge parametrization (see Eq. \req{e6}) the growth equation becomes
($n_z=1/\sqrt g$):
\begin{eqnarray}
\partial_t h &=& {\sqrt g}{\cal G}_{z} \nonumber\\
&=&
\sqrt g \left[  \unabla \left( \frac{ \unabla h}{\sqrt g}\right)
\left(n_z\frac{d\chi(n_z) }{dn_z} - \chi(n_z)\right)
+ \frac{d^2 \chi(n_z) }{d{n_z}^2} 
\frac{\U{\nabla} h\cdot\U{\nabla}}{g}
\frac{1}{\sqrt{g}}\right] \nonumber \\
&=& 
(\chi_1'-\chi_1) [ \nabla^2 h + \frac{1}{2} \nabla^2 h ( \unabla h)^2]
+\frac{1}{2}(\chi_1-\chi_1'-\chi_1'') 
\unabla( \unabla h (\unabla h)^2)
+\ldots
\label{neweq}
\end{eqnarray}
where the last line contains a small gradient expansion with 
$\chi_1\equiv \chi(1), \chi_1'\equiv d \chi(x)/dx|_{x=1}$ etc.
Note that even though 
the first two terms in Eq. \req{neweq} are the same as those in 
Eq. \req{srfmg} for the surface tension, the latter equation is rotationally
invariant while the former is not, due to the presence of the other
terms in Eq. \req{neweq}.   
Eq. \req{neweq} also implies that the linear relaxation of the
interface ({\em via} the term $\nabla^2 h$) is related to the 
behavior of $\chi(n_z)$ near $n_z=1$. From our previous 
discussion, one expects that the effect of the constraint in RSOS models
is negligible on flat regions, i.e., $\chi_1\simeq 0$. It is also reasonable 
to assume that $\chi(n_z)$ is an increasing function of $n_z$ 
which attains its maximum at $n_z=1$. This would indeed 
favour flat regions over inclined ones. 
This would therefore lead to a positive coefficient
$\chi_1'-\chi_1 >0$ in Eq. \req{neweq}.

In principle one cannot exclude situations in which 
$\chi_1'-\chi_1 <0$. This would lead to a linear instability in the
growth equation of the same form as the one which has been
invoked by Mazor {\it et al.}, (1988) to describe the columnar morphology 
of thin films (see later). 

%The constraint discussed above
%would result in a restoring force toward a flat surface. This is accomplished
%if $\chi_1' > \chi_1$ and it is satisfied, e.g., by the choice
%$\chi_1 (n_z) = {\nu}_z {n_z}^2 $ corresponding to a maximum for $n_z=1$
%(${n_z}^2 \leq 1 $). 
It is interesting to note that in $D=1$ the two 
nonlinear terms in Eq. \req{neweq} have the same form 
($=(dh/dx)^2 d^2 h / dx^2 $) and the coefficient 
$\chi_1-\chi_1'-\case3/2\chi_1''$ may assume both positive and 
negative values while keeping $\chi_1'-\chi_1 >0$.

Here and in the  following, if not stated
otherwise, we  assume  that the direction  $\hat{z}$ is the
same as that  along which  $h$ is  measured. The  case in which a
different direction is singled out can be dealt with similarly.
A specific example will be discussed in the next section.

\subsubsection{The flux of particles and geometric effects}
\label{three.A.6}

Apart from forces that can be derived from an Hamiltonian, 
effective forces  resulting from purely   geometrical effects  
may also appear in Eq. \req{eq1}. The simplest such non--conservative
term  derives from  a flux  of   particles, whose
velocity is  $\vec{\Phi}$,  that reaches  the  surface and sticks to
it. The  external flux  of incoming
particles  is  the  basic  source of  non--conservative  noise in
interface   growth. For  this reason, we  consider  here only the
effect of the  average flux  $\vec{J}=\avg{\vec{\Phi}}$ while the
fluctuation term $\vec{F}=\vec{\Phi}-\vec{J}$ will be the subject
of section \ref{three.B}. The  growth rate ${\cal G}_J$ produced by
$\vec{J}$ is  proportional to  the flux of  $\vec{J}$ through the
surface,  and is exactly  a measure  of how many  particles have
been added to the surface:
\begin{equation}
{\cal G}_J=-\hat{n}\cdot\vec{J}=J n_J,
\label{gflux}
\end{equation} 
where  $n_J$ is the  component of $\hat{n}$ in the
$-\vec{J}$ direction,  $J=|\vec{J}|$ and the negative sign recalls
that  $\hat{n}$  and  $\vec{J}$ have   opposite  directions. 
When
growth occurs from a  vapor, the average of $\vec{\Phi}$ 
at the position $\vec r$ of the interface is proportional to the
normal $\hat n$ at $\vec r$, i.e., $\langle \vec \Phi \rangle= - 
\lambda\hat n$,
implying ${\cal G}_{\rm vap}=\lambda$. Thus, growth in the 
presence of vapor has
the same effect as the pressure term in subsection 2.
On the contrary, when particles arrive at the surface in a collimated 
beam this term can be eliminated by a Galilean transformation.
Indeed, in the framework of Monge parametrization one gets:
\begin{equation}
\left.\partial_t h\right|_{\rm flux}=\sqrt{g}{\cal
G}_J=-J_z+\U{J}_\bot\cdot\vec{\nabla}h,
\label{gfluxmg}
\end{equation} 
with  $\U{J}_\bot$ being the projection of $\vec{J}$
on the substrate  plane. The first term is  absorbed into $\partial_t h$
by the transformation $h\to  h-J_z t$ while the second disappears,
once $\U{x}\to \U{x}+\U{J}_\bot t$.

Let us return to the height constraint used
in some discrete lattice models of the growth of interfaces.
Apart from producing a surface tension term, the constraint also has
an effect on the flux term. 
Let  us consider a vertical flux
$\vec{J}\parallel\hat{z}$ for simplicity ($\U{J}_\bot =0$). 
At a microscopic scale, deposition can only occur in local minima of 
the discretized surface. 
In a coarse grained picture, minima are rare on  steep portions 
of the interface and deposition would be less probable there.
In other words, the constraint should reduce the flux 
$d\phi=n_Jd^Ds\sqrt{g}$ through the infinitesimal surface 
element $d^Ds\sqrt{g}$, if $n_J\equiv n_z$ is
small. This is easily accounted for by multiplying Eq.
\req{gflux} by an increasing function of $n_z,\Upsilon(n_z)$
\begin{equation}
{\cal G}_J=J n_z~\Upsilon(n_z).
\label{gfluxKPZ}
\end{equation}    
In the small gradient  expansion in $h$,  this term produces the
nonlinear term  of the KPZ  equation for any choice of the  function
$\Upsilon(x)$, provided it is  increasing.
Indeed, if $\vec{J}\parallel\hat{z}$, 
one has 
\[\partial_t h|_{flux}=J \sqrt g n_z \Upsilon(n_z) = 
J \Upsilon (1- (\unabla h)^2/2+ \ldots)\]
which, in a Taylor expansion,
contains the non--linear term of  the KPZ equation with a negative
$\lambda$. This result was also found by other 
means (Krug, 1989; Tang {\it et al.}, 1992). 
The effect of a constraint on height differences thus produces
both the surface tension, as seen in the previous section, 
and the nonlinear term of the KPZ equation.

Of course the applicability of Eq. \req{gfluxKPZ} extends  to any
situation where the probability for an incoming particle to stick
on the surface  depends on the local  inclination of the surface.
This can  also  account for the  relation  between  the ballistic
deposition  model and the KPZ  equation. In  this model 
(Meakin {\it et al.}, 1986), particles
come from a beam downwards in a straight line and 
attach  at the  first site they reach in
their trajectories which has a nearest neighbor surface site. The
result of this  mechanism is a  non--compact  cluster with a fixed
density. Even  though overhangs  are present in  the surface, the
description of the  process in terms of a  single valued function
$h(\U{x},t)$ is  possible, at a mesoscopic  level, by considering
$h(\U{x},t)$ as the $z$  coordinate of the highest occupied
site for each  $\U{x}$. An incoming  particle may either stick at
the surface or  penetrate into the voids of the structure. In the
latter case, the deposition process will not result in an increase
of  $h(\U{x},t)$.  Particles that arrive on a flat
portion of the  interface have a higher probability of 
penetrating into the
structure than those arriving on steep ones. This situation would
also be modeled by Eq. \req{gfluxKPZ}, with a function 
$\Upsilon(n_z)$ that, in  contrast to the previous
case, should now be a decreasing  function of  $n_z$; 
in the small gradient expansion,
this would finally result in a non--linear  term of the  
KPZ type  with  a positive coefficient $\lambda$.

The flux term produces another  interesting effect in situations
where  the  atoms  cannot be   approximated as  point  particles.
Mazor {\it et al.}, (1988) have shown  that the finite
size of  particles plays an important role in MBE
experiments  in  which  thin  films  are grown  at  intermediate
temperatures. The basic  observation is that, if the atoms have a
radius $\xi$, deposition does  not actually  occur on the surface
but at a  distance  $\xi$ from  it in the  normal  direction. The
growth rate is then  proportional to the flux of the beam through
a surface $\vec{r}\ '(\vec{s}) =\vec{r}(\U{s})+\xi \hat{n}$ that is
displaced by an  amount $\xi$, in the  normal direction, from the
actual     surface.  The   growth  rate  on a    surface  element
$d\sigma=d^Ds\sqrt{g}$         is     proportional  to  the  flux
$d\phi=-\hat{n}'\cdot\vec{J} \sqrt{g'} d^Ds$   of $\vec{J}$ through
the  surface  $\vec{r}\ '$. Here  the primes refer to the displaced
surface. Since   $\hat{n}'=\hat{n}$, the only  effect comes through the
$\sqrt{g}$ factor. The metric tensor ${g'}_{ij}$ is 
obtained by observing that
${\partial_i}\vec{r}\ '={\partial_i}\vec{r}+\xi{\partial_i}\hat{n}$
so that ${g'}_{ij}={\partial_i}\vec{r}\ '\cdot\partial_j\vec{r}\ '=
g_{ij}-2\xi b_{i, j}+O(\xi^2)$ (see Appendix \ref{appA}, 
especially Eq. \req{sfqf}).
Evaluating the  determinant of ${g'}_{ij}$, we find that Eq.
\req{gflux} has to be modified to:
\begin{eqnarray}
{\cal G}_{J'}&=&{\cal G}_J+{\cal
G}_\xi=\frac{d\phi}{d\sigma} \nonumber \\
&=&-\hat{n}\cdot\vec{J}\sqrt{\frac{g'}{g}}=
-\hat{n}\cdot\vec{J}(1-\xi H) + O(\xi^2).
\label{xi}
\end{eqnarray}
In the equation for $h(\U{x},t)$ a term:
\begin{equation}
\sqrt{g}{\cal G}_\xi=-\nu_\xi\U{\nabla}\frac{\U{\nabla} 
h}{\sqrt{g}}= -\nu_\xi\nabla^2 h+\ldots
\label{Gximg}
\end{equation}  
must  be  included in the  growth  equation, with
$\vec{J}\parallel \hat{z}$ and $\nu_\xi=J\xi$ which is positive. 
Thus, the finite size of the incident particles gives rise to an effective
anti-diffusive behavior in the growth equation.

An interesting extension of the results of this section 
arises when considering a beam of particles which is not 
perpendicular to the substrate. This setting has also been
considered by Meakin (1988b) and Krug and Meakin (1989,1991)
for ballistic deposition processes. From our perspective, one 
generally expects that, in this situation, a coupling between 
the interface fluctuations and the flux term arises. 
Let us consider, for example, how an oblique
flux modifies the effect of the non-zero radius of particles just
discussed. It is straightforward to derive
the small gradient expansion of Eq. \req{xi} in the case $\vec{J}= 
(\U{J} _{\bot}, J_{z})$, where the z direction 
is normal to the substrate:
\begin{equation}
\sqrt{g}{\cal G}_\xi= J_{z} -
\U{J}_{\bot} \cdot \vec{\nabla} h - \xi J_{z} \nabla ^{2} h +
\xi\U{J}_{\bot} \cdot \U{\nabla} h \nabla ^{2} h +\ldots
\label{oblieq}
\end{equation}
As already mentioned, the first two terms are eliminated by an
appropriate choice of the reference frame. The third term in Eq.
\req{oblieq} has just been discussed. The last one  is new
and it represents the coupling of interface fluctuations with the
transverse component of the oblique flux. It has been  argued
by Marsili {\it et al.}, (1996), applying
dynamical renormalization group techniques (see, e.g., 
Ma and Mazenko, 1975; Forster {\it et al.}, 1977; Frei and T\"auber 1994)
that this term, in competition with surface diffusion, 
is responsible for a new scaling behavior of 
interface fluctuations at intermediate scales.
This approach predicts a roughness exponent 
$\alpha=1/3$ which
is in very good agreement with experiments of MBE, where
a value $\alpha=0.30 - 0.33$ was measured (Salvarezza {\it et al.}, 1992;
Herrasti {\it et al.}, 1992).

One can, in general, imagine other effects arising from an 
inclined flux. For example, if $\U{J} _{\bot}\ne 0$ a new
term would also arise from the effect of a constraint on height
differences on the flux term described in Eq. \req{gfluxKPZ}.
Indeed, if $\hat{n}\cdot\vec{J}= n_z + 
\U{n}_\bot\cdot\U{J}_\bot$, one would  expect  a term 
$(\U{J}_\bot\cdot\U{\nabla}h )(\U{\nabla} h)^2$ in the gradient
expansion. Common wisdom ( i.e., power counting) however, suggests
that this term will not modify the leading scaling behavior of
interface fluctuations in the presence of the KPZ non--linearity
$(\U{\nabla} h)^2$.

\subsubsection{Surface diffusion}
\label{three.A.7}

In cases in which the binding energy of particles on the surface is
large compared to  thermal energy  fluctuations, the motion of the
particles is constrained  to be along the surface.
This is  actually the case in  many  experimental realizations of
MBE. A force acting in a direction normal to the surface cannot 
displace the particles. Its effect is to 
change the local chemical potential $\mu$.  Differences in
chemical   potential on  the  surface, in turn, produce a
current  proportional  to  the  gradient  of  $\mu$  on the
surface (Villain 1991).  The evolution of  the surface is governed by
the continuity equation relating the particle density $\rho$ and this
current: $\partial_t\rho\propto\hat{n}\cdot\partial_t\vec{r}$ 
yields the
left--hand side of Eq. \req{eq1} while the divergence of the current gives
the right--hand side.

The mathematical translation of this argument is as follows.
The constraint that the motion of particles occurs on the surface
implies that the volume ${\cal V}$ it encloses cannot change. Since
\begin{equation}
\partial_t{\cal V}=\int d^Ds\frac{\delta {\cal V}}
{\delta\vec{r}(\U{s},t)}\partial_t\vec{r}(\U{s},t)
=\int d^Ds\sqrt{g}\hat{n}\cdot\left[\hat{n}{\cal G}+
\vec{F}(\U{x},t)\right],
\label{dvdt}
\end{equation}
the condition $\partial_t{\cal V}=0$ for the deterministic part of Eq.
\req{eq1} implies:
\begin{equation}
\partial_t{\cal V}=0~~~\Leftrightarrow~~~\int d^Ds\sqrt{g}{\cal G}=0.
\label{gcons}
\end{equation}
A sufficient condition for this to hold is: 
\begin{equation}
{\cal G}=-\bl{\cal F}=-\U{\rm div}\ \U{{\cal J}}.
\label{Gcons2}
\end{equation}
In other words volume conservation requires ${\cal G}$ to be the
(covariant) divergence of a (contravariant) current ${\cal J}^i$
( see Eq. \ref{diveq}).
Here $\U{{\cal J}}$ is the surface current of particles and
${\cal F}$ is proportional to the chemical potential. This
relaxation mechanism is known as surface diffusion.
If this force, $\cal F$, derives from a potential, then:
\begin{equation}
\hat{n}\cdot {\cal F}=-\frac{1}{\sqrt{g}}\frac{\delta {\cal H}}{\delta \vec{r}}.
\label{Gcons3}
\end{equation}    
Having derived,  in the previous section,
the forces coming  from the simpler contributions to the 
potential energy of an
interface, it is  straightforward to find the appropriate
${\cal G}^c$ term in the equation for $\partial_t \vec{r}$ under surface
diffusion by applying the  Beltrami Laplace operator to the terms
derived  previously.  In the  Monge form,  the term  appearing in
Eq. \req{mongeq1} has the form:
\begin{equation}
\sqrt{g}{\cal G}^c=-\sqrt{g}\bl{\cal F}=\sqrt{g}\bl\frac{\delta 
{\cal H}}{\delta h}
\label{consmg}
\end{equation}
where the second equality holds for growth mechanisms that
can be derived from a potential (using the result of
Appendix \ref{appB}).
When surface diffusion occurs to
minimize the surface area, the corresponding term in Eq.
\req{eq1} reads:
\begin{equation}
{\cal G}_s^c=-\mu_s\bl\left(\hat{n}\cdot\bl\vec{r}\right)=- \mu_s\bl H.
\label{surfc}
\end{equation}
This follows from the two previous equations and Eq. \req{surf}.
This term has been widely used in numerical and analytical
studies (Mazor {\it et al.}, 1988; Golubovi\'{c} and Karunasiri, 1991; 
Siegert and Plischke, 1992; Sun and Plischke, 1993; Sun and Plischke, 1994). 
Its expression, in terms
of $h(\U{x},t)$, is
\begin{equation}
\sqrt{g}{\cal G}_s^c=-\mu_s \sqrt g \bl\U{\nabla}\cdot\frac{\U{\nabla} 
h}{\sqrt{g}} \simeq -\mu_s (\nabla^2)^2 h + \ldots
\label{srfcmg}
\end{equation}

A pressure  term does not produce any surface
diffusion, since surface diffusion entails the conservation 
of the volume.
External potentials, that depend on $\vec{r}$,
can produce surface diffusion. In the simplest case of
the gravitational potential (along the $\hat{z}$ direction) we have:
\begin{equation}
{\cal G}_g^c=\nu_g~\hat{z}\cdot\hat{n}~ H
\label{gravc}
\end{equation}
where     $\nu_g=\rho   a_g$ is   the  mass   density   times the
gravitational acceleration.    The explicit
$\hat{n}\cdot\hat{z}$ factor indicates that this term breaks
rotational invariance. ${\cal G}_g^c$ has the same form as ${\cal
G}_\xi$ in Eqs. \req{xi} and \req{Gximg}, but  has the opposite  sign. 
In actual  experimental situations however, the  gravitational  
energy is  negligible with respect to the binding energy. 
It can be  estimated, for example, that in MBE
growth of thin films  $\nu_g$ is of the order of
$10^{-14}$ of the corresponding coefficient in the ${\cal G}_\xi$
term. However, terms of the form:
\begin{equation}
\sqrt{g}{\cal G}_g^c=\nu_g \sqrt{g} \bl h=
\nu_g \U{\nabla}\cdot
\frac{\U{\nabla} h}{\sqrt{g}}
\simeq \nu_g \nabla^2 h + \ldots
\label{gravmg}
\end{equation}
with a positive  $\nu_g$ have often been 
used in recent publications (see, e.g., Golubovi\'{c} and Karunasiri, 1991). 
This is  principally justified by renormalization group
considerations. Even if this  term  is  not present at a
microscopic  level, it is  generated  in the  iteration of the renormalization
group equations  from the non--linear terms.

Surface  diffusion  can also be induced by a curvature dependent
Hamiltonian or  by orientation  dependent  potentials in the same
manner. The explicit  expression is simply  given by that for the
non--conserved case with the additional Beltrami--Laplace operator.

\subsection{Stochastic Evolution}
\label{three.B}

As already mentioned, the principal source of randomness in
interface growth comes from a flux of particles that deposit on
the surface. Another kind of noise is produced by thermal
fluctuations of the surface. The main difference is that 
the latter conserves the total volume enclosed by the surface. 
As discussed previously, the random force 
$\vec{F}=\hat{n}\eta$ is in the
normal direction and we can take $\avg{\eta}=0$.
In some representations,e.g., if one studies the growth of a
spherical substrate in polar coordinates, this choice cannot be made.
The properties of $\eta$ will now be discussed for the two
different cases of non-conservative and conservative noise. Our
main concern is the correlator $\avg{\eta(\vec{s},t)\eta(\vec{s}\ ',
t')}$. In cases when the statistics of the noise is Gaussian, as
is almost always assumed, this correlator specifies the entire
distribution of $\eta(\vec{s},t)$. (Non-Gaussian statistics for the
noise, especially with long tails, has been shown by Zhang (1990)
and Krug (1991) to affect the scaling properties of the interface).

\subsubsection{Non-conservative noise}
\label{three.B.1}

For a flux  of particles  arriving at  the surface  with velocity
$\vec{\Phi}$,  the  noise term  is given  by  $\eta=\hat{n} \cdot
\vec{F}$ where
$\vec{F}=\vec{\Phi}-\vec{J}$, with $\vec{J}=\avg{\vec{\Phi}}$.
It describes the
fluctuation in the number of particles arriving at the
surface.    The     correlator  of   $\eta$  is  then   given by:
\begin{eqnarray}
\avg{\eta(\vec{s},t)\eta(\vec{s}\ ',t')}&=&n_\alpha(\vec{s},t) 
n_\beta(\vec{s}\ ',t') \avg{F^\alpha(\vec{s},t) 
F^\beta(\vec{s}\ ',t')}\nonumber\\
&=& n_\alpha(\vec{s},t) n_\beta(\vec{s}\ ',t') \Gamma^{\alpha
\beta}\frac{\delta(\vec{s}-\vec{s}\ ')}{\sqrt{g}}\delta(t-t').
\label{ncn1}
\end{eqnarray}
Here we choose $\vec{F}$ to be delta correlated both in
space and time. The more general choice:
\[\avg{F^\alpha(\vec{s},t) F^\beta(
\vec{s}\ ',t')}= \Gamma^{\alpha\beta} C(\vec{s}-\vec{s}\ ',t-t'),\]
where $C(\U{s},t)$ is a R--invariant function, involves no
further complication, so we will use the explicit expression
\req{ncn1}. The coefficients $\Gamma^{\alpha\beta}$ are
symmetric in the indices and specify the geometric properties of
the noise. We may distinguish two extreme cases:
\begin{itemize}
\item deposition processes that occur from the condensation of
an isotropic vapor and
\item deposition that occurs from a collimated beam of particles.
\end{itemize}
The difference between these two possibilities fully appears in
the Monge parametrization.
In order to discuss this, it is useful to
introduce the random field:
\begin{equation}
\tilde{\eta}(\U{x},t)=\sqrt{g}\eta(\U{x},t),
\label{noisemg}
\end{equation}  
which is the quantity that  appears  explicitly in the
equation    for   $\partial_t  h$.  The     correlation    properties of
$\tilde{\eta}(\U{x},t)$    are  derived   directly from  those of
$\eta(\U{x},t)$ discussed above.

>From equation \req{ncn1} one easily sees that:
\begin{equation}
\avg{\tilde{\eta}(\U{x},t)\tilde{\eta}(\U{x}',t')}=
\frac{\Gamma^{zz}-2{\partial_i} h 
\Gamma^{iz}+{\partial_i} h \partial_j h \Gamma^{ij}}{\sqrt{g}}
\delta(\U{x}-\U{x}')\delta ( t-t').
\label{tetacor}
\end{equation}
If growth occurs from the condensation of an isotropic vapor,
we expect that $\vec{F}$ is a random
vector with uncorrelated components and\footnote{Here we use the 
Kronecker symbol with both upper indices which coincides with the metric
tensor $g^{\alpha\beta}$ in the $D+1$ dimensional space.}
$\Gamma^{\alpha\beta}=\Gamma \delta^{\alpha\beta}$.
In equation \req{tetacor} we find:
\begin{equation}
\avg{\tilde{\eta}(\U{x},t)\tilde{\eta}(\U{x}',t')}=
\Gamma\sqrt{g}\delta (\U{x}-\U{x}')\delta (t-t').
\label{vapmg}
\end{equation} 

For growth occurring from a directed flux, one may assume that
all components of $\vec{F}$ are independent random
variables so that $\Gamma^{\alpha,
\beta}=\Gamma_\alpha\delta^{\alpha\beta}$ (no  summation on
$\alpha$ is assumed  here). If rotational  invariance is expected
for rotations in the substrate plane we have:
$\Gamma_i=\Gamma_\parallel$ and 
\begin{equation}
\avg{\tilde{\eta}(\U{x},t)\tilde{\eta}(\U{x}',t')}=
\frac{\Gamma_z+\Gamma_\parallel(\U{\nabla}h)^2}{\sqrt{g}}
\delta (\U{x}-\U{x}')\:\delta (t-t').
\end{equation}
Note that in--plane correlations are enhanced in regions
where $h$ has steep derivatives.
For a collimated beam perpendicular to the interface we have
$\Gamma_\parallel\ll\Gamma_z$ and 
\begin{equation}
\avg{\tilde{\eta}(\U{x},t)\tilde{\eta}(\U{x}',t')}\cong
\frac{\Gamma_z}{\sqrt{g}}\delta (\U{x}-\U{x}')\:\delta (t-t'),
\label{vertbeam}
\end{equation}
while if $\Gamma_\parallel\cong\Gamma_z$ we recover Eq. \req{vapmg}.
If the randomness only affects
the intensity of the beam,  $\Gamma^{\alpha\beta}=\Gamma 
J^\alpha J^\beta$ and equation \req{tetacor} becomes:
\begin{equation}
\avg{\tilde{\eta}(\U{x},t)\tilde{\eta}(\U{x}',t')}=
\Gamma\frac{\left(J_z-\U{\nabla}h\cdot\U{J}_\perp\right)^2}
{\sqrt{g}}\delta (\U{x}-\U{x}')\:\delta (t-t')
\label{tempo}
\end{equation}
where $\U{J}_\perp$ is the component of $\vec{J}$ in the substrate
plane $\vec{r}=\U{x}$. In the case of  vertical rain, $J_z\gg
|\U{J}_\perp|$ and we recover the previous result Eq. \req{vertbeam}.

The  physical meaning of the prefactors  of the  delta functions is
evident if we introduce $\eta_o(\U{x},t)$  such  that
$\avg{\eta_o(\U{x},t)\eta_o(\U{x}',t')}=\Gamma \delta
(\U{x}-\U{x}')\delta  (t-t')$. In  condensation from a vapor, Eq.
\req{vapmg},      we   find   that         $\tilde{\eta}(\U{x},t)
=g^{1/4}\eta_o(\U{x},t)$    in Eq.   \req{mongeq1} (Since $g$ contains
stochastic variables, this is also the correct mathematical interpretation of
the correlation in Eq. \req{vapmg}).  The noise is
enhanced in  regions  where $h$ has  steep  derivatives since the
exposed  surface area in  the  substrate element  $d^Dx$
is larger by a factor of $\sqrt{g}$. The  opposite case  is that of growth
from a   perpendicular beam   $\vec{J}=J_z\hat{z}$ in  which case
$\tilde{\eta}(\U{x},t)=g^{-1/4}\eta_o(\U{x},t)$.       This is
because the flux of $\vec{J}$ through the surface is proportional
to $\hat{n}\cdot\hat{z}=1/\!\sqrt{g}$ and  regions  with high
slopes receive less particles than those that are flatter.

\subsubsection{Conservative noise}
\label{three.B.2}

Another source of noise comes from thermal fluctuations and from
internal degrees of freedom of the interface. In this case the
noise is called conservative because it causes no increase of
the volume enclosed by the interface. Using Eq. \req{dvdt}, 
this requirement is
translated into the condition:
\begin{equation}
\left.\partial_t{\cal V}\right|_{noise}=
\int d^Ds\sqrt{g}\eta=0
\label{dvdtnoise}
\end{equation}
where again we have taken $\vec{F}=\hat{n}\eta$.
This poses a condition on $\eta$. A  general
way to let the noise contribution in Eq. \req{dvdtnoise} vanish
is to take:
\[\eta=\U{\rm div}\ \U{\zeta},\] where
$\U{\rm div}$ is the covariant divergence acting on the vector $\U\zeta$
that is a delta correlated noise both in space and time:
\[\avg{\zeta_i(\U{s},t)\zeta^j(\U{s}',t')}=\Gamma\delta_i^{\; j}
\frac{\delta(\U{s}-\U{s'})}{\sqrt{g'}}\delta(t-t').\]
Here reparametrization invariance has been satisfied and the
delta function allows the use of $g'=g(s')$ instead of
$g(s)$. The correlations of $\eta$ readily follows:
\begin{eqnarray}
\avg{\eta(\U{s},t)\eta(\U{s}',t')}
&=&\frac{1}{\sqrt{gg'}}{\partial_i} 
\left\{ \sqrt{g}\partial_j' \left[ \sqrt{g'}
\avg{\zeta^i(\U{s},t)\zeta^j(\U{s}',t')}\right]\right\} \nonumber\\
&=&\Gamma\frac{1}{\sqrt{g g'}}
{\partial_i} \left[ \sqrt{g} g^{ij}\partial_j'
\delta(\U{s}-\U{s'})\delta(t-t') \right] \nonumber\\
&=&-\Gamma\frac{1}{\sqrt{g}}
\left[{\partial_i}\sqrt{g} g^{ij}\partial_j
\frac{\delta(\U{s}-\U{s'})}{\sqrt{g'}}
\delta(t-t')\right]\nonumber\\
&=&-\Gamma\bl\frac{\delta(\U{s}-\U{s'})}{\sqrt{g'}}\delta(t-t')
\label{cn1}
\end{eqnarray} 
where the primed  quantities refer to $\U{s}'$ and the
presence of the delta function has been used repeatedly to change
from primed to unprimed  quantities (note that the operator $\bl$
does not  act on  $g'$). This  is the  natural  generalization in
reparametrization invariant  form of the correlator often used in
dealing with conserved noise that contains the Laplacian operator
acting on the delta function. The expression of the correlator in
the Monge   parametrization is  readily  derived  from the above
expression.

\subsubsection{Approach to equilibrium}
\label{three.B.3}
In previous sections we have frequently dealt with terms in the 
deterministic part of the growth equation of the form:
\beq
\frac{\partial}{\partial t}h(\U{x},t)=-\sqrt{g}
\Gamma\frac{\delta {\cal H}}
{\delta h(\U{x},t)}+\tilde\eta(\U{x},t),
\label{langevin}
\eeq
where $\Gamma=1$ for non-conservative dynamics, and 
$\Gamma=-\bl$ for conservative dynamics.
If the noise is Gaussian with correlation (see Eq. \req{vapmg}): 
\beq
\avg{\tilde\eta(\U{x},t)\tilde\eta(\U{x'},t')}= 2 T\sqrt{g}\Gamma
\delta(\U{x}-\U{x}')\delta(t-t'),
\label{eqcorr}
\eeq
one can write the associated Fokker--Planck equation in the form:
\beq
\frac{\partial}{\partial t}P[h(\U{x}),t]=\int d^D x
\sqrt{g}\Gamma \frac{\delta}{\delta h(\U{x},t)}
 \left[T \frac{\delta}{\delta h(\U{x},t)}
+\frac{\delta {\cal H}}{\delta h(\U{x},t)}
\right]~P[h(\U{x}),t].
\label{fpeq}
\eeq

Eq. \req{fpeq} holds if a  suitable regularization has been chosen such that
$\frac{\delta}{\delta h(\U{x},t)}$ commutes with $\sqrt{g}\Gamma$. This is
fulfilled, e. g., in the dimensional regularization scheme commonly used
in field theoretical treatments (see, e.g., Zinn-Justin, 1993).

In Eq. \req{fpeq} $P[h(\U{x}),t]$ is the probability 
functional which yields the probability of the
interface configuration $h(\U{x})$ at time $t$.
It is easy to see that the stationary distribution, obtained
by setting the right--hand side to zero, is given by 
\beq
P[h(\U{x}),t\to\infty]\propto\exp\left\{-\frac{{\cal H}[h(\U{x})]}
{T}\right\}.
\label{steady}
\eeq
The equality of the coefficient in front of the functional 
derivative and of that in front of the noise correlation is
often referred to as the fluctuation dissipation theorem (see, e.g, Deker and
Haake, 1975).
When this theorem holds the stationary distribution is given
by Eq. \req{steady}.
It is interesting to note that the R--invariant form of the functional 
derivative, with the R--invariant form of the correlator deriving from an
isotropic growth mechanism ($\Gamma=1$) yields a dynamics
which leads to the equilibrium distribution Eq. \req{steady}.
This was first noted by Bausch {\it et al.}, (1981).
One may apply these considerations to the R--invariant form of
the KPZ equation describing growth from condensation of a vapor
as discussed by Maritan {\it et al.}, (1992). 
However,  as
already pointed out, ${\cal H}_{KPZ}$ of Eq. \req{hkpz} is unbounded 
as $h\to\infty$ 
and therefore Eq. \req{steady}, not being normalizable, is
meaningless.

Note also that if the flux term breaks rotational invariance,
as in the case of a collimated vertical beam (see Eq. \req{tempo}), 
the fluctuation
dissipation relation does not hold even if the deterministic
part derives from a potential. 

Furthermore we note that, for $\Gamma=-\bl$, the deterministic
part has the same form discussed in section \ref{three.A.7} and the 
correlation of the noise is exactly that derived above in Eq. 
\req{cn1}. Therefore we can conclude that a deterministic conservative 
dynamics with a conservative noise leads to the stationary state 
described by Eq. \req{steady} (provided $P[h,t]$ is normalizable).

\section{Discussion}
\label{three.C}

We have  seen that there  are four  different mechanisms 
producing  a  Laplacian term $\nu\nabla^2 h$ 
in  the small  gradient  expansion of
Eq.\req{mongeq1}. Three of them, the  surface tension,  an orientation 
dependent potential (or a constraint on
$n_z$) and surface diffusion induced  by  gravity,  lead to a
positive $\nu$  coefficient that drives the evolution towards
flatter and  flatter surfaces.  Note that one can distinguish
between these  effects 
only through higher order terms in the  gradient
expansion, which may however be irrelevant in the renormalization
group sense. The fourth
mechanism, related to the finite  size of the aggregating
particles, gives a negative contribution to the coefficient 
in front of the Laplacian, so it would produce an   instability
if acting alone.  This is   evident  since Eq.
\req{Gximg}   can formally be   derived  as a  surface  diffusion
induced  by a  {\em  negative}   gravitational  field 
(${\nu}_g = - {\nu}_\xi$ in Eq. \req{gravmg}).  
This also
implies  that  this term strictly
conserves the   volume   enclosed  by the   surface   (while 
surface tension, Eq.
\req{srfmg}, does  not) even  though it was not  derived from
conservation considerations.

Secondly we  note that  the non--linear term of  the KPZ equation
\req{kpz}, can be derived in one of three ways: from a pressure
term in  a  potential,  that  gives a  positive   $\lambda$ for a
growing surface, from growth due to condensation of a vapor or  
from an inclination  dependent factor in the
flux term. The latter may result from the effect of a constraint on
$n_z$,  yielding a  negative  $\lambda$, in  agreement with known
results on restricted solid on solid models (Meakin, 1993). 
It has also been  argued that such a term is expected in
ballistic     deposition but  with  the   opposite   effect, i.e.,
inhibiting  growth on  flat portions  of the  surface. A positive
$\lambda$ is  expected in this  case. The  change of sign in this
coefficient   does  not  change  the   character of  the  process
dramatically as it does for $\nu$. The value of $\lambda$ however
is a directly measurable  quantity (Krug, 1989) since it is related
to  the   inclination   dependence  of  the average velocity of 
growth.  
In this way Krug (1989) was able to predict
the  presence  of the  non--linear term  of the  KPZ  equation in
various  models. Note that
this is a  criterion  based on the global  behavior, while our analysis
is based  on the local properties of the growth process. Another
derivation of the KPZ equation for restricted solid on solid 
models is based on
its relation with the directed polymer problem (Kardar and Zhang, 1987, 
Fisher and Huse, 1991)in a random
environment as shown, e.g., by Tang {\it et al.}, (1992).

Equation \req{curv2mg} provides  a further physical derivation of
a term $(\nabla^2)^2 h$ in the  equation for a growing interface.
This term  has usually been associated with  surface diffusion.
This extends the validity of  the results derived in the presence
of this term to  situations where the  restoring force is derived
from a potential corresponding to the surface energy being 
proportional to $H^2$. 
The expansion of the potential energy of the interface in powers
of $H$ will, in general, contain also a constant term $H^0$, 
proportional to the surface area. Straightforward analysis shows
that in the presence of this term all higher powers are 
irrelevant from a renormalization group 
point of view. There are situations, as 
in the dynamics of fluid membranes, where this term is known
to be absent. The leading terms would then result from the terms
linear and quadratic in $H$.
An interesting point for future investigation is an analysis
of the effects of the term resulting from a potential linear in $H$,
Eq. \req{curv1mg}.

Sun {\it et al.}, (1989) have
studied an equation with a linear term proportional to
$(\nabla^2)^2 h$ and a non--linear term proportional to $\nabla^2
(\U{\nabla} h)^2$. The same equation was studied by 
Lai and Das Sarma (1991) and Kim and Das Sarma (1994) (see
also Das Sarma {\it et al.}, 1996).
This is in a loose sense the conserved version
of the KPZ equation.
We note  here that  while the  first term  could be  derived from
surface diffusion or from a  potential proportional to $H^2$, the
non--linear  one is not  related to one  of the  simple mechanisms
discussed here. In particular, a derivation of the {\em conserved}
KPZ equation  cannot follow the same lines  described above. In
one case, the KPZ  equation comes  out as a  result of a
process that does  not conserve the  volume. In the second, it 
derives
from the effect of a  constraint on height  gradients and one 
needs to motivate the rather odd choice
$\Upsilon(n_z)=-\bl\tilde{\Upsilon}(n_z)$. 
There are no R--invariant potentials that would lead to such a
term in the gradient expansion, either for conservative or 
non--conservative dynamics. It is still possible however that such
a term is generated dynamically in a renormalization
group procedure.

Note also that in the noise term, there are prefactors  
in front of the random field. In a small gradient expansion these 
yield a nonlinear term of the KPZ
type with a  random valued  $\lambda$.
It is also important to note that the occurrence of multiplicative 
noise could  significantly  change
the scaling properties of growing interfaces. 

We also discussed the generalization of the fluctuation dissipation 
theorem to the R--invariant form of the growth equation. We recovered
the observation of Bausch {\it et al.}, (1981) that  R--invariance and 
rotational invariance are enough to establish fluctuation
dissipation relations for non--conserved dynamics. We also
found that this observation generalizes  to conserved 
dynamics (both the noise and the deterministic part being conservative).

Recently, Keblinski, Maritan, Toigo, Koplik and Banavar (1994) and
Keblinski {\it et al.}, (1996) 
have introduced a simple 
continuum model that allows
for overhangs and an arbitrary topology of the growing interface. 
The model captures surface diffusion in a natural manner and,
with an appropriate
aggregation mechanism, it produces growth normal to the interface.
The model equation consists of two parts, the first is  conserved order
parameter dynamics that allows for the definition of a topologically 
unrestricted interface and builds the correct physics of surface
diffusion, whereas the second term provides the growth and roughening
at the interface.

In the simplest version, their equations are:

\begin{equation}
\frac{\partial f(\vec{r},t)}{\partial t} = \Gamma \nabla ^{2} \frac{\delta
F}{\delta f(\vec{r},t)} + I, 
\label{f1eq}
\end{equation}
 
\noindent
where
\beqar
F &=& \int \left [ - \frac{1}{2} f^{2} + \frac{f^{4}}{4} + a(\nabla f)^{2}
\right ] dv,
\label{ffeq}\\
\hbox{and}\qquad  
I&=&C_1\vert \vec \nabla f \vert + D_1 \sqrt {\vert \vec \nabla f \vert} 
\eta(r,t),
\label{I1eq}
\eeqar
where $\eta$ is a Gaussian noise  uncorrelated in time and space with a
width equal to 1 and mean value 0.

Equation (\ref {f1eq}) without the $I$ term has a simple interpretation -- 
it is merely the deterministic part of the standard 
model - B dynamics (Hohenberg and Halperin, 1977) that conserves the 
order parameter.

The choice of the sign of the coefficient of $f^2$ in the expansion for the 
free energy
(\ref{ffeq}) corresponds to a temperature lower than $T_c$,
 so that  the two values $\pm 1$
of the order parameter $f(\vec r,t)$
minimizing
the free energy describe the two equilibrium phases of the system.

An interface can  naturally be
defined as the crossover region between the $f=-1$ and $f=+1$ regions - and, 
operationally, a point
$\vec r_i$ is defined to be on the interface when $f(\vec r_i,t)=0.$ 

$a>0$ is the surface diffusion coefficient and
sets the intrinsic length  in the system.  Indeed
the  width of the interface is proportional to $\sqrt{a}$. 
Also  the effective strength of the surface tension turns out to be 
proportional to $\sqrt{a}$. 

The $I$ term
allows for the growth and fluctuation of the interface. 
The $\vec \nabla f$ factor ensures that the growth and 
fluctuations are operative only in the vicinity of the 
interface -- away from the interface $\vec \nabla f$ is effectively zero.
The positive coefficients $C_1$ and $D_1$ are  the magnitudes of the growth
and noise respectively. 
The $\vec \nabla f$ factor  produces growth
 normal to the interface. 

Numerical results (Keblinski, Maritan, Toigo, Koplik and Banavar, 1994;
Keblinski {\it et al.} 1994, 1995, 1996) 
show that this model 
is in the same universality class as the KPZ equation.
The $I$ growth mechanism not only gives growth normal to the interface
but also the rate of the growth per unit length is constant along the
interface, equal to $C_1\int_{s_{min}}^{s_{max}}\vert \vec \nabla f \vert ds
= 2C_1$ where the integral is performed across
the interface and $s_{min}$  and $s_{max}$ are defined by
$ f(s_{min})=-1$ and $f(s_{max})=1$
(note that the integral is independent of the shape of the $\vec \nabla f$ 
profile as long
as $f$ increases monotonically from $f=-1$ to $f=+1$).
 This feature leads to KPZ-like behavior. 

The model presented above has a local conservation law to avoid the 
formation of islands 
with $f=-1$ in the region predominantly with $f=+1$, and vice-versa.
However, the conserved and non-conserved model should 
exhibit interfaces with the same behavior. 
It can be shown (Keblinski {\it et al.}, 1996)
that the non-conserved version of this model 
is equivalent to Eq. \req{MKPZ} with a noise term of the form \req{noisemg} with
variance given by \req{vapmg}  when the interface is sharp, i.e., $a$ 
is small.
Thus, this model may be interpreted as a continuum version of the 
Eden growth model, with 
redistribution of the aggregated particles via surface diffusion.
The Eden (1958) model  is known to be in the KPZ universality class -- 
the surface
diffusion in the limit of large length scales does not change the 
geometrical properties of the interface, but introduces a short range smoothing
mechanism. 

The growth of 
real surfaces is often influenced by non-local effects such as 
screening or shadowing. When the aggregating particles 
follow linear trajectories one can expect that, if the roughness
is large enough,
some parts of the interface are 
shadowed and, therefore, do not grow (Tang and Liang, 1993). 
In order to accommodate  this
phenomenon, the previous model can be extended (Keblinski {\it et al.},
1995, 1996) 
to incorporate the dynamics of the
depositing vapor and non-local effects. 

The extended model involves two fields $f$ and $\phi$ and is 
governed by the equations:

\begin{eqnarray}
\frac{\partial f(\vec{r},t)}{\partial t}&=& \nabla ^{2} \frac{\delta
F}{\delta f(\vec{r},t)} + B (\nabla f)^{2} \phi(\vec{r},t)\nonumber\\
& & +C\sqrt{(\nabla f)^{2} \phi} \eta(\vec{r},t)
\label{feq}
\end{eqnarray}
\begin{equation}
\frac{\partial \phi(r,t)}{\partial t} = \vec{\nabla} [D\vec{\nabla}
\phi(\vec{r} t) - \vec{A} \phi(\vec{r},t)] - B(\nabla f)^{2} \phi(\vec{r},t)
\label{geq}
\end{equation}
 
\noindent
with $F$ again given by  Eq. (\ref{ffeq}).
While the first part of Eq. (\ref{feq}) is identical
to that  described earlier,  the growth mechanism
is different. Now the growth of the $f$ field  occurs at the expense of
the $\phi$ field. The $\phi$ field represents the local density of the incoming
particles towards the interface, and Eq. (\ref {geq}) describes the
dynamics of the depositing vapor. The first part of Eq. (\ref{geq})
 is simply the diffusion
equation in the presence of an external force $\vec A$. In order to analyze
the growth arising from ballistic trajectories, $D$
has to be chosen much smaller
than $\vec A$, so the $\vec A\phi$ flux is the primary mechanism for $\phi$ 
field transport\footnote{Here $D$ is not to be confused with the 
spatial dimension.}. The aggregation relies
on the conversion of the $\phi$ field into the $f$ field as described
by the coupling term $B$ in Eqs. \req{feq} and \req{geq}. The $\nabla f$
 factor in the $B$ term  makes 
the aggregation
operative only within an interfacial "skin" region of the aggregate 
with its width proportional to $\sqrt{a}$. The $\phi$ factor in the aggregation 
term
ensures that the  growth occurs only
if $\phi>0$. The B term acts as a sink for the diffusive field $\phi$, and its 
magnitude
is chosen to be sufficiently large to convert  all  of the $\phi$  into
 $f$  within the interfacial region  effectively leading to $\phi \sim 0$ below
the interface ($f \sim +1$).
 Shadowing effects are naturally incorporated in the equations. When
the $\phi$ field trajectory intercepts the skin, the $\phi$ field is
 converted into $f$ and any subsequent interception occurs with
$\phi=0$ and, therefore,  does not lead to the growth of  the shadowed part
of the interface. Note that  non-local effects are incorporated
in a local way in Eqs. \req{feq} and \req{geq}. One does not need to monitor
the geometry of the interface to incorporate shadowing -- it is implemented 
dynamically by
the  $\phi$ field.
 The conversion of the $\phi$ field
into the $f$ field at the vicinity of the interface does not depend crucially on
the particular functional form of the coupling chosen. The rate of growth is
effectively equal to the  intensity of the incoming $\phi$ field flux.
In addition to the growth term,  there is a fluctuation
term  $C$  in Eq. (\ref {feq}).
The Gaussian $\eta(\vec r,t)$
 factor is the same as was introduced  previously. 
In this manner, the fluctuations in the strength of the incoming
$\phi$ flux are incorporated, since the aggregation rate  
is equal to the  intensity of the  incoming $\phi$ flux.

Strikingly, the model presented in Eqs. (\ref {feq}) and (\ref {geq})
can be straightforwardly modified (Keblinski {\it et al.}, 1994, 1995, 1996)
to model diffusion--limited--aggregation type of phenomena 
(Witten and Sander, 1981; Niemeyer {\it et al}. 1984;
see also Pietronero and Tosatti, 1985).
One just needs to substitute the  ballistic flux with a diffusive flux
that is responsible for the transport of the aggregating field.

The equations are modified to:

\begin{equation}
\frac{\partial f(\vec{r},t)}{\partial t} = \nabla ^{2} \frac{\delta
F}{\delta f(\vec{r},t)} + J(\vec r, t),
\label{f2eq}
\end{equation}
\begin{equation}
\frac{\partial \phi(r,t)}{\partial t} = D\vec{\nabla}^2 
\phi(\vec{r}, t) - J(\vec r, t), 
\label{g2eq}
\end{equation}
\noindent
with
\begin{equation}
J(\vec r, t) = - \vec \nabla f \cdot D \vec \nabla \phi \cdot \eta(\vec r, t),
\label{J1eq}
\end{equation}
\noindent
where $\eta (\vec r,t)$ is a Gaussian noise   with 
non-zero average value $V>0$ and width W.

The first part of  Eq. (\ref {f2eq}) and
the free energy $F$ is the same as described 
previously. The interaction term $J>0$, leads to the growth of $f$ and 
decay of $\phi$ such that $f + \phi$ is a conserved quantity  and changes only
due to the sources of the $\phi$ field at the boundary. 
The R--invariant form of the growth equation
provides an interesting alternative approach to avoid the
no-overhang approximation, even though it cannot describe
non-local effects like shadowing in ballistic aggregation
and screening in the diffusion limited aggregation models.

The inadequacy of the no--overhang approximation
is just one of the reasons why one may want to  rely on 
the R--invariant formulation of growth problems. 
For example, the R--invariant form of the equation displays 
the full invariance properties with respect to space
translations and rotations and the conservation laws
which the process satisfies. These are lost, for any but 
infinitesimal transformations, once one restricts
attention to the lowest order terms in the gradient 
expansion. These invariances play a crucial role in
the implementation of renormalization group approaches
around the lower critical dimension (see, e.g., Bausch {\it et al.}, (1981),
because they provide conditions for the renormalizability
of the theory. 

A second situation in which a full R--invariant form of the growth
equation would be preferable to the lowest order gradient
expansion arises when the scaling behavior is determined
by a strong coupling fixed point.
In order to appreciate this situation it is preferable to 
sketch briefly the standard approach of the perturbative 
dynamical renormalization
group (Ma and Mazenko, 1975). Let us consider the Langevin equation
\[\partial_t h(\U{x},t)=\hat{\cal L}\,h(\U{x},t)
+\lambda\hat{\cal N}[h(\U{x},t)]+
\eta(\U{x},t)\]
where $\hat{\cal L}$ is a linear
differential operator in the $\U{x}$ variable
and $\hat{\cal N}$ is a non--linear combination of 
$h$ and gradients. As an example, in the KPZ equation \req{kpz}
one has $\hat{\cal L}= \nabla^2 $ and $\hat{\cal N}[h]=(\U{\nabla}h)^2$.
The noise is Gaussian with zero mean and
$\langle \eta(\U{x},t)\eta(\U{x}',t)\rangle =
2\Gamma\delta^{D}(\U{x}-\U{x}')\delta(t-t')$.
If $\lambda=0$, the equation can easily be solved in Fourier 
space. It is therefore easy to find the exponents $\alpha_0$ and
$z_0$ which characterize the scaling in the linear theory:
Under a rescaling of length by a factor of $\ell$, the
time will scale by a factor $\ell^{z_0}$ and $h$ will acquire
a factor $\ell^{\alpha_0}$. One can then analyze  the
effect of a small non--linearity ($\lambda\ll 1$) on the dynamics.
If the above discussed change of scale in the linear theory
affects the non--linearity with a factor $\ell^y$ ( i.e., if
$\hat{\cal N}[h]\to \ell^y\hat{\cal N}[h]$ as $\U{x}\to\ell\U{x}$,
$t\to\ell^{z_0} t$ and $h\to\ell^{\alpha_0} h$)
and $y<0$, one can conclude that the nonlinearity is irrelevant.
On the other hand, if $y>0$ one concludes that the effect of the 
non--linearity will increase as the scale increases and
will eventually dominate the large scale behavior of interface
fluctuations. It usually happens that $y$ is a decreasing 
function of $D$, the dimensionality of the substrate, and there 
is a dimension $D_c$ below which the non--linearity is relevant.
In this case, the full program of the dynamic renormalization group
becomes necessary. Since the method can in principle treat 
only small non--linearities, being based on a perturbation expansion,
it usually provides estimates of the exponents close to the
dimension $D_c$, where one can assume that since $y$ is small, the 
fixed point is accessible within the expansion.
If this procedure works, i.e., if one finds a stable
fixed point whose distance from the fixed point of the
linear theory is small when $D_c - D$ is small, one can also
conclude that all higher order terms, which have been neglected
in the gradient expansion, are irrelevant (if $\alpha_0<1$).
This whole program, which is a very powerful tool to estimate
critical exponents, fails if one finds no stable fixed point
or if the non--linearity turns out to be relevant only if $\lambda$ is bigger
than a critical value $\lambda_c$. The nature of the phase
for $\lambda>\lambda_c$ turns out to be outside the range
of perturbative methods and in this situation,
one has no reason  to neglect the higher
order terms in the gradient expansion. 

This situation is realized in the most studied model of 
growth, namely the KPZ equation, for which above the substrate dimension
$D_c=2$ (which is also the physical dimension in which
one is interested) the non-linearity becomes relevant only
if $\lambda>\lambda_c$. 

There is a general consensus that different models for non-equilibrium
dynamics, such as  restricted solid on solid models (Meakin, 1993),
the Eden model (Eden, 1958) and directed polymers in random
media (Halpin-Healy and Zhang, 1995), fall in the same universality class.
Such an expectation is mainly based on the observation that
the small gradient expansion of these models contains
the terms in the KPZ equation \req{kpz}. 
While analytical and numerical results in $D=1$ almost unambiguously 
support this expectation, there is no reason, in principle, to 
believe that in $D\ge 2$ the strong coupling regime
of these models is described by the same exponents. In fact,
as we have shown, the continuum equations for the Eden model, for RSOS models
and for ballistic aggregation differ in the higher order 
terms of the gradient expansion. A yet different gradient 
expansion can be obtained for the free energy of
directed polymers. Going one step further than the
usual KPZ truncation, Marsili and Bray (1996) studied the
equation
\beq
\frac{\partial h}{\partial t}=\nu\nabla^2h 
+\kappa(\vec{\nabla}h)^2\nabla^2 h +
\lambda + \frac{\lambda}{2} (\vec{\nabla}h)^2 +\ldots 
+ \eta.
\label{modkpz}
\eeq
This derives from the small gradient expansion in the case of
RSOS growth, as described in section \ref{three.A.5}. This equation 
indeed coincides with Eq. \req{neweq} with $\chi_1=0$,
$\chi_1'=-\chi_1''>0$. 
Note that here {\em both} surface diffusion {\em and} the
growth term have been expanded to the same (second) order.
It turns out that a mean field, infinite dimensional limit of Eq. 
\req{modkpz} can be meaningful only for $\kappa>0$
(Marsili and Bray, 1996).
This analysis reveals that 
steps or bumps of a finite height develop on the surface and that 
dynamical scaling (Family and Vicsek, 1991) is not satisfied.
These results, which differ substantially from those obtained 
in the same limit for directed polymers by Derrida and Spohn (1988)
(see also Cook and Derrida, 1991), raise doubts on the existence
of a single universality class for all these processes. 
In high dimension, the $\kappa$ term turns out to be necessary
in order to avoid finite time singularities which occurs in the
simpler KPZ equation ($\kappa=0$). These finite time singularities
turn even worse in the case of the Eden model or  ballistic
aggregation. Indeed, the model Eq. \req{modkpz} can be 
generalized to these process with $\kappa<0$ which 
describes the suppression of surface tension on steep portions
of the interface discussed by Maritan {\it et al.}, (1992).
These instabilities suggest that the Monge representation
of the interface may not be adequate to describe these 
processes (Maritan {\it et al.}, 1992).

Anomalous dynamic scaling has also been reported ( Das Sarma {\it et al.}, 
1996) for various models proposed to describe MBE. More precisely, 
numerical results show that the local scaling properties, defined 
in terms of the correlation function, identify an exponent $\alpha_{\rm loc}$
which happens to be different from the one which describes 
the behavior of the global surface thickness $W(L,t)$ with the size
$L$ of the sample at saturation. This anomalous behavior has been 
related by Schroeder et al. (1993), Das Sarma et al. (1994) and
Krug (1994) to the peculiar statistics of steps ( i.e., height
differences between neighboring points). The broadness of this 
distribution (which diverges for infinite times and infinite $L$)
again raises doubts on the validity of the small gradient expansion.

In all these situations, we believe the concept of reparametrization
invariance may prove to be an invaluable starting point 
for elucidating the correct physics.

\section{Acknowledgments}

We are indebted to L\'aszl\'o Barab\'asi for helpful correspondence.
This work was supported by grants from EPSRC, INFM, NASA, NATO, NSF 
and the Donors of the Petroleum Research Fund administered by the American 
Chemical Society.

\appendix

\section{Differential Geometry}
\label{appA}

An orthonormal basis is assumed in $D+1$ dimensional space and Greek
letters are  used  for the   vector   components.
Latin  letters used as an index  refer to the
components of vectors in  the  $D$   dimensional
parametrization   space. $s^i$ are general  curvilinear coordinates
that label
points on the $D$ dimensional surface.  The  notation
${\partial_i}=\partial/\partial  s^i$ is  used for  covariant derivatives
\footnote{We have avoided introducing covariant derivatives in order 
to maintain the exposition as simple as possible. Thus $\partial_i$
and $\partial^i$ introduced here behave like tensors only when
applied to scalar quantities.}.
Summation over repeated indices is always assumed. Lastly, for the
scalar  product  in  both  spaces a  dot is  used while  $\times$
denotes the vector product.

The distance between infinitesimally close points on the surface
is given by the first fundamental quadratic form:
\begin{equation}
|d\vec{r}|^2={\partial_i}\vec{r}ds^i\cdot
\partial_j\vec{r}ds^j=g_{ij}ds^i ds^j.
\label{ffqf}
\end{equation}

This defines the  metric tensor:
$g_{ij}={\partial_i}\vec{r}\cdot\partial_j\vec{r}$.
$g=\det\{g_{ij}\}$  denotes  its determinant  while $g^{ij}$ is
the inverse: \footnote{The symbol
$\delta$ is used here for the Kronecker  delta. It will also be used
for the Dirac delta function and for functional differentiation.} 
$g_{ik}g^{kj}=\delta_i^{\; j}$.
The metric tensor and its inverse are also used in the usual way 
to lower and raise indices, i.e, $v_i=g_{ij} v^j$ or $v^i=g^{ij} v_j$.
We also use the notation $\partial^i=g^{ij} \partial_j$.

The only  restriction on the choice
of the  parametrization is that $g\ne 0$,  i.e., that $g_{ij}$ is
invertible, and  this implies  also that ${\partial_i}\vec{r}\ne 0$. 
The vectors ${\partial_i}\vec{r}$  lie  in the  tangent  
hyperplane  so that  the normal versor  is given  by  
$\hat{n}=    g^{-1/2}\partial_1\vec{r} \times
\partial_2\vec{r}   \times  \ldots  \times   \partial_D\vec{r}$, 
where $g^{-1/2}$ ensures normalization 
($ {\vec v}_1 \times \ldots \times {\vec v}_D \equiv
{\varepsilon}_{{\alpha}_0{\alpha}_1\ldots {\alpha}_D} 
v_{1{\alpha}_1} \ldots v_{D{\alpha}_D}$, 
where
${\varepsilon}_{{\alpha}_0{\alpha}_1\ldots {\alpha}_D}$ is the completely
antisymmetric Levi--Civita tensor and is equal to $(-1)^P$ with
$P$ being the order of permutation of 
${{\alpha}_0{\alpha}_1\ldots {\alpha}_D} $ with respect to $1,2, \ldots,D+1$.)

A quantity ${T_{i \ldots}}^{j \ldots}(\U{s})$ is said to be a tensor if, under
the change of parametrization $\U{s}'(\U{s})$, it transforms as:
\[{{T}'_{i \ldots}}^{j \ldots}(\U{s}')=
\frac{\partial s^k}{\partial {s'}{^i}} \ldots 
\frac{\partial {s'}{^j}}{\partial s^\ell} {{T}_{k \ldots}}^{l \ldots}(\U{s}).\]
>From Eq. \req{ffqf} one sees that $g_{ij}$ and its inverse $g^{ij}$
are tensors.

A quantity $\varphi(\U{s})$ is said to be a 
scalar if
$\varphi'(\U{s}')=\varphi(\U{s})$ .
In particular $r_\alpha(\U{s})$ and $n_\alpha(\U{s})$, with 
$\alpha=1,2, \ldots, D+1$ are scalar quantities, while
$\partial_i \varphi$ and $\partial^i \varphi$ are particular
cases of tensors called co--vector and vector respectively.

The invariant surface element is given by $d\sigma=d^Ds\sqrt{g}$
and this implies that the invariant form of the delta function in
parameter space is:
\begin{equation}
\delta_0(\U{s}-\U{s}')=\frac{\delta(\U{s}-\U{s}')}{\sqrt{g}},
\label{RIdelta}
\end{equation}
where $\delta(\U{s})$ is the usual delta function in $D$
dimensional space.  Thus
$\int d\sigma f(s) \delta_0(\U{s}-\U{s}')=f(\U{s}')$

For differential calculus, invariant forms of the gradient,
divergence and curl are obtained requiring the transformation
properties of tensors to apply. The gradient of a scalar $S$ is
simply given by ${\partial_i} S$ while the divergence of a vector 
is\footnote{Indeed, if $\varphi$ and $v^i$ are a scalar and a contravariant 
field respectively, Eq. \req{diveq} follows on requiring 
$\int d^D s \sqrt{g} v^i\partial_i\varphi=
-\int d^D s \sqrt{g} \varphi\U{\rm {div}}\  \U{v}$.}:
\beq
\U{\rm {div}}\  \U{v}=\frac{1}{\sqrt{g}} \partial_i(\sqrt{g} v^i).
\label{diveq}
\eeq

Taking the divergence of the contravariant gradient
yields the reparametrization invariant generalization of the
Laplacian operator in curved space:
\begin{equation}
\bl=\frac{1}{\sqrt{g}}{\partial_i}\sqrt{g}\partial^j
=\frac{1}{\sqrt{g}}{\partial_i}\left(\sqrt{g}g^{ij}\partial_j\right)
\label{bldef}
\end{equation}
which is known as the Beltrami--Laplace operator.\footnote{Eq. \req{bldef}
can easily be deduced by requiring 
$\int d\sigma \partial_i\varphi \partial_j\varphi g^{ij} =
-\int d\sigma \varphi\bl\varphi$ for any scalar field $\varphi$.}

The curvature $\kappa$ of the surface along a curve $\U{s}(\ell)$
is given by 
$\hat{n}\cdot \frac{{\partial}^2\vec{r}(\U{s}(\ell))}{\partial\ell^2}$
( $\ell$ is the arc length).
Since $\hat n \cdot \partial_i\vec r = 0 $, $\kappa$ may be written in terms of
the second fundamental quadratic form:
\begin{equation}
\kappa= b_{ij}\frac{ds^i}{d\ell} \frac{ds^j}{d\ell},
\label{sfqf}
\end{equation}
where $b_{ij}=\hat n \cdot \partial_i\partial_j \vec r = - \partial_i\hat n
\cdot \partial_j \vec r$.\footnote{The first form shows that $b$
is a symmetric matrix, whereas the second shows that it is a tensor:
hence the R--invariance of its eigenvalues.}
This defines the principal curvatures (directions) as the
eigenvalues $\lambda_i$ (vectors) of $b_i^j$. These are invariant
under reparametrization. The mean curvature $H$ is the sum of
these and thus equals \footnote{Actually
the mean curvature should contain a factor $1/D$ that
we disregard for convenience.} the trace of $b_i^j$:
\begin{equation}
H=b_i^i=\sum_{i=1}^D\lambda_i=-{\partial_i}\hat{n}\cdot{\partial^i}\vec{r}.
\label{Hdef}
\end{equation}

Another useful definition of $H$ comes from observing that, since
$\hat{n}\cdot{\partial^i}\vec{r}=0$,  
${\partial_i} \left( \sqrt{g}\hat{n}\cdot{\partial^i}\vec{r} \right) =
\sqrt{g}({\partial_i}\hat{n})\cdot{\partial^i}\vec{r}+
\hat{n}\cdot{\partial_i} \left(\sqrt{g}{\partial^i}\vec{r} \right) =0$. 
This implies:
\begin{equation}
H=-{\partial_i}\hat{n}\cdot{\partial^i}\vec{r}=\hat{n}\cdot\bl\vec{r}.
\label{Hdef2}
\end{equation}

The Gaussian curvature is 
defined as $K=\det\{b_i^j\}=\prod_i\lambda_i$.
Furthermore it is easy to see that
\beq
\partial_i \hat n = - b_{ij} \partial^j \vec r.
\label{didn}
\eeq
Indeed $\partial_i \hat n \bot \hat n $ and 
${\partial_i} (\hat{n}\cdot{\partial_j}\vec{r})=0$ implies 
${\partial_i} \hat{n}\cdot{\partial_j}\vec{r}=-\hat n \cdot 
\partial_i\partial_j \vec r=-b_{ij}$.

\subsection{The Monge Form}
\label{appA.1}

A particular choice of parametrization is the Monge form:
\begin{equation}
\vec{r}=\left(\U{x},h(\U{x})\right),
\label{mongedef}
\end{equation}
where $\U{x}$ is a vector in the $D$
dimensional substrate plane and $h(\U{x})$ is the height of the surface in
the direction $\hat{z}$ perpendicular to this plane. Use of this
parametrization implies that no overhangs are present in the surface
since otherwise $h(\U{x})$ would not be single valued. 
In this parametrization the metric tensor has the form: 
\beq
g_{ij}=\delta_{ij}+{\partial_i} h\partial_j h \qquad {\rm and}
\qquad g^{ij}=\delta_{ij}-\frac{1}{g} {\partial_i} h\partial_j h,
\label{metric}
\eeq
where:
\begin{equation}
g=1+(\U{\nabla}h)^2\hbox{ and }\qquad ; \qquad 
\hat{n}=\frac{1}{\sqrt{g}}(-\U{\nabla}h,1)
\label{gn}
\end{equation}
and
\beq
{b^i}_j=g^{ik}b_{kj}=\partial_i \frac{\partial_j h}{\sqrt g}.
\label{bdef}
\eeq
Finally the mean curvature is given by
\begin{equation}
H={b^i}_i=\U{\nabla}\frac{\U{\nabla} h(\U{x})}{\sqrt{g}}.
\label{curvmongedef}
\end{equation}

The equation for $h(\U{x},t) $ is obtained from \req{eq1} considering the
various components of $\vec{r}$. 
On defining $\vec{r}(\U{s},t)=\left(\U{x}(\U{s},t),
h(\U{s},t)\right)$, we get:
\begin{eqnarray}
\partial_t h(\U{s},t)&=&n^z{\cal G},\nonumber\\
\partial_t x^i(\U{s},t)&=&n^i{\cal G},\nonumber
\end{eqnarray}
where $n^i$ and $n^z$ are the components of the normal in
the directions $\hat{x}^i$ and
$\hat{z}$ respectively. These derivatives are evaluated at
constant $\U{s}$ while we are
interested in the derivative of $h$ at constant $\U{x}$, 
\beq
\partial_t h(\U{s},t)=\partial_t h(\U{x},t)+\frac{\partial h}{\partial x^i}
\partial_t x^i(\U{s},t).
\label{mongelast}
\eeq
where $h(\U{x},t) \equiv h(s(\U{x},t),t)$.
>From the above equations and $\hat{n}=(-\U{\nabla}h,1)/\!\sqrt{g}$,
one readily finds the deterministic part of Eq. \req{mongeq1}.

\section{Equations derived from a potential in the 
Monge representation}
\label{appB}

The property that the functional derivative 
$\frac{\delta {\cal H}}{\delta \vec r}$ is orthogonal to the vector
${\partial_i}\vec{r}$ translates into:\[\frac{\delta {\cal H}}{\delta h}
{\partial_i} h + \frac{\delta {\cal H}}{\delta x^i}=0.\]
This allows one to eliminate the functional derivative w.r.t. $x^i$ in
\[-\sqrt{g}{\cal G}=\hat{n}\cdot\frac{\delta {\cal H}}{\delta \vec{r}} 
=n^z \frac{\delta {\cal H}}{\delta h} + n^i\frac{\delta {\cal H}}{\delta x^i}=
\sqrt g \frac{\delta {\cal H}}{\delta h}, \]
and to find Eq. \req{dhmonge}.

\section{Derivation of the Growth Term Due to Surface Energy}
\label{appC}

In the functional derivative of $g$ w.r.t. $\vec{r}$, in Eq.
\req{srf}, we use the
fact that $g$ is a determinant and the property:
\beq
\delta \ln\det\hat{M}=\delta{\rm tr}\ln \hat{M}={\rm tr}
\hat{M}^{-1}\delta \hat{M}
\label{logdet}
\eeq
that holds for variations of a matrix $\hat{M}$.
This allows one to write:
\[\delta\int d^Ds\sqrt{g}=\frac{1}{2}\int d^Ds\sqrt{g} g^{ij}\delta g_{ji}
=-\int d^Ds{\partial_i}
(\sqrt{g} g^{ij}\partial_j\vec{r}) \cdot\delta\vec{r}
=-\int d^Ds
\sqrt{g} \bl \vec r \cdot\delta\vec{r} \]
for a variation $\delta\vec{r}$ of $\vec{r}$, which readily yields:
\begin{equation}
-\frac{1}{\sqrt{g(\U{s})}}\frac{\delta}{\delta\vec{r}(\U{s})}\int
d^Ds'\sqrt{g(\U{s}')}=\frac{1}{\sqrt{g(\U{s})}}{\partial_i}
\left(\sqrt{g(\U{s})} 
g^{ij}(\U{s})\partial_j\right)\vec{r}(\U{s}) =\bl \vec{r}(\U{s}),
\label{csrf}
\eeq
where Eq. \req{bldef} has been used.

\section{ Curvature Dependent Potential }
\label{appD}

For a variation $\delta\vec{r}$ in $\vec{r}$, $\hat{n}$ changes to
$\hat{n}+\delta\hat{n}$. Since $\delta(\hat{n}\cdot\hat{n})=0$,
the variation $\delta\hat{n}$ is normal to $\hat{n}$. Consider Eq.
\req{Hdef2} for $H$. In
\[\delta H=\delta\hat{n}\cdot\bl\vec{r}+\hat{n}\cdot\delta(
\bl\vec{r})\]
the first term on the right--hand side vanishes since $\bl\vec{r}$ is
parallel to $\hat{n}$. The variation of $\bl\vec{r}$, by simple
arithmetic, is:
\begin{eqnarray*}
\delta\bl\vec{r}&=&\delta\frac{1}{\sqrt{g}}{\partial_i}\left(\sqrt{g}g^{i,
j}\partial_j\right)\vec{r}\\
&=&\left({\partial_i}\frac{\delta\sqrt{g}}{\sqrt{g}}\right)
{\partial^i}\vec{r}+\bl\delta\vec{r}+\frac{1}{\sqrt{g}}
{\partial_i}\left[\sqrt{g}(\delta
g^{ij})\partial_j\right]\vec{r}.
\end{eqnarray*}
The first term in the last line, being proportional to
${\partial^i}\vec{r}$, vanishes once a scalar product with 
$\hat{n}$ is taken. For the
same reason, the only contribution that survives in the last term
is obtained when the derivative ${\partial_i}$ acts on $\vec{r}$. The
variation $\delta g^{ij}$ is expressed in terms of $\delta\vec{r}$
by taking the variation of $g^{ik}g_{k,j}=\delta^i_{\;j}$ so that
finally:
\[\delta H =\hat{n}\cdot \bl \delta \vec{r}-2
\left({\partial^i}\delta\vec{r}{\partial^j}\vec{r}\right)
\left(\hat{n}\cdot{\partial_i}\partial_j\vec{r}\right).\]
The variation of ${\cal H}_{c,1}=\kappa_1\int d^Ds\sqrt{g} H$
w.r.t. $\delta\vec{r}$ involves the variation $\delta H$ and the
variation of $\sqrt{g}$, which is evaluated as before. The
functional derivative of the first term is evaluated from the
above equation with a partial integration:
\[\frac{1}{\sqrt{g}}\frac{\delta {\cal H}_{c,1}}{\delta\vec{r}}=
k_1\left\{-H\bl\vec{r}-{\partial^i}\vec{r}
{\partial_i} H+\bl\hat{n}+2\frac{1}{\sqrt{g}}{\partial_i}
\left[\sqrt{g}(\hat{n}\cdot{\partial^i}
\partial_j\vec{r}){\partial^j}\vec{r}\right]\right\}.\]
On multiplying the above equation
by $\hat{n}$ to find ${\cal G}_{c,1}$, the property
$\hat{n}\cdot{\partial_i}\vec{r}=0$ can be used again to show that the
second term gives no contribution and the last term 
becomes $2b^i_{\; j}b^j_{\; i}$.
This is twice the trace of the square of the matrix of the coefficients of the
second fundamental form. Finally we have to compute
$\hat{n}\cdot\bl\hat{n}$. Using the facts that
$\hat{n}\perp{\partial^i}\hat{n}$ and
that ${\partial_i}\hat{n}=-b_i^{\; j}
\partial_j\vec{r}$, we easily find:
\begin{equation}
\hat{n}\cdot\bl\hat{n}=-{\partial_i}
\hat{n}\cdot{\partial^i}\hat{n}=-b_i^{\; j}b_j^{\; i}=
-\sum_{i=1}^D \lambda_i^2.
\label{nbln}
\end{equation}
Collecting the various terms we get the result displayed in Eq.
\req{curv1} that is clearly fully R--invariant.

The chain rule of differentiation, applied to
${\cal H}_{c,p}=\kappa_p\int d^Ds\sqrt{g} H^p$,
also gives Eq. \req{curvp}, where the second term, as before
comes from the variation of $\sqrt{g}$ while the others come
from $\delta H^p=p H^{p-1}\delta H$. This also needs 
the
straightforward generalization of Eq. \req{nbln} to:
\[\hat{n}\cdot\bl(F\hat{n})=\left(\bl-\sum_{i=1}^D \lambda_i^2
\right)F\]
for a generic R--invariant function $F(\U{s})$.

\section{ Orientational Energy }
\label{appE}

The variation of \req{hnz} for a change $\vec r\to\vec r+\delta \vec r$ can 
be obtained as soon as we know $\delta\sqrt g$ and $\delta \hat n$.
The former has already been obtained in appendix C and it is 
$\delta\sqrt g=\sqrt g\ g_{ij} \partial_i \vec r \cdot\partial_j 
\delta\vec r$ whereas
the latter is derived as follows. Since $\delta\hat n\perp\hat n$ and 

\[0=\delta(\hat n\cdot\partial_i \vec r)=\partial_i\vec r\cdot\delta
\hat n +\hat n\cdot\partial_i\delta\vec r, \] 
one has:
\beq
\delta\hat n = -\partial^i\vec r (\hat n\cdot\partial_i\delta\vec r).
\label{e1}
\eeq
The variation of Eq. \req{hnz} is then:
\beqar
\delta {\cal H}_z &=&\int d^D s\sqrt g [-g^{ij}(\partial_i\vec r 
\cdot\partial_j
   \delta\vec r) \chi(n_z)  \nonumber\\
  &+&\frac{d\chi(n_z)}{ dn_z} (\hat z\cdot\partial^i\vec r)\ (\hat n\cdot
   \partial_i\delta\vec r)] \, \label{e2}
\eeqar
from which it follows that:
\beqar
\frac{1}{\sqrt g} \frac{\delta {\cal H}_z}{\delta\vec r} 
&=&\frac{1}{\sqrt g}\partial_j
\Big[\sqrt g g^{ij} \partial_i\vec r \chi(n_z)- \sqrt g \frac{d\chi(n_z)}{dn_z}
\hat n (\hat z\cdot\partial^j\vec r)\Big]\nonumber\\
&=& \chi (n_z)\bl\vec r -\frac{d^2\chi (n_z)}{d n^2_z} \hat n (\hat z\cdot
\partial^i\vec r) \partial_i n_z   
- \frac{d\chi(n_z)}{d n_z} \hat n (\hat z \cdot\bl \vec r) \nonumber\\
&-& 
\frac{d\chi (n_z)}{d n_z} (\partial^j\vec r\cdot\partial_j n_z 
-(\hat z\cdot \partial^j\vec r)\partial_j \hat n) \ . 
\label{e3}
\eeqar

The last two terms sum to zero since $\partial_i\hat{n}=
-b_i^j\partial_j \vec{r}$, and due to the symmetry of $b_{ij}$. 
Finally, using Eq. \req{Hdef2} ( i.e., $\bl\vec r =\hat n H)$ we get:
\beq
-\frac{1}{\sqrt g} \frac {\delta {\cal H}_z}{\delta\vec r} = \hat n G_z,
\label{e4}
\eeq
with
\beq
G_z= H \Big(-\chi(n_z) +n_z \frac{d\chi(n_z)}{dn_z} \Big) + \frac{d^2\chi(n_z)}
{dn^2_z}  \partial^i z\partial_i  n_z,
\label{e5}
\eeq
where $z\equiv \hat z\cdot \vec r$.

Using Monge parametrization,
\beq
G_z = H \left(n_z \frac {d\chi(n_z)}{\d n_z} -\chi(n_z) \right) + 
\frac{d^2\chi (n_z)}{d n^2_z} \frac {\partial_i h \partial_i g^{-1/2}}{g},
\label{e6}
\eeq
with $n_z=g^{-1/2}$ and $H = \vec\nabla \Big( \frac{\vec\nabla h}{\sqrt g}
\Big)$.


\begin{thebibliography}{99}

\bibitem[ ]{ref} Amar, J. G., and F. Family, \cpre{47}{1595}{1993}

%\bibitem[ ]{ref} Amit, D. J., 1978, {\it Field theory,the
%renormalization group and critical phenomena} (World Scientific,
%Singapore).

%\bibitem[ ]{ref} Bales G. S., and Zangwill A., \cprl{63}{692}{1989}.

\bibitem[ ]{ref} Bales, G. S., R. Bruinsma, E. A. Eklund, R. P. U.
Karunasiri, J. Rudnick, A. Zangwill, 1990, Science {\bf 249}, 264.

\bibitem[ ]{ref} Barab\'asi, L., and H. E. Stanley, 1995, {\it Fractal
Concepts in Surface Growth} (Cambridge University Press).

\bibitem[ ]{ref} Bausch, R., V. Dohm, H. K. Janssen  and R. K. P. Zia,
\cprl{47}{1837}{1981}.

\bibitem[ ]{ref} Bruinsma, R., and G.  Aeppli, \cprl{52}{1547}{1984}.

\bibitem[ ]{ref} Buldyrev, S. V., A.-L. Barab\'asi, F. Casetra, S.
Havlin, H. E. Stanley and T. Vicsek, \cpra{44}{R8313}{1992}.

\bibitem[ ]{ref} Cieplak, M., and M. O. Robbins, \cprl{60}{2042}{1988}.

\bibitem[ ]{ref} Cook, J., and B. Derrida,\jsp{63}{505}{1991}.


\bibitem[ ]{ref} Das Sarma, S., 1994, Fractals {\bf 1}, 784.

\bibitem[ ]{ref} Das Sarma, S., C. I. Lanczycki, R. Kotlyar and S. V.
Ghaisas, \cpre{53}{359}{1996}.

\bibitem[ ]{ref} Das Sarma, S., and P. Tamborenea, \cprl{66}{325}{1991}.

\bibitem[ ]{ref} Das Sarma, S., and R. Kotlyar, \cpre{50}{R4275}{1994}.

\bibitem[ ] {ref} Das Sarma, S., S. V. Ghaisas and J. M. Kim,
\cpre {49} {122} {1994}.

\bibitem[ ]{ref} Das Sarma, S., and S. V. Ghaisas, \cprl{69}{3762}{1992}.

\bibitem[ ]{ref} Deker, U., and F. Haake, \cpra{11} {2043} {1975}.

\bibitem[ ]{ref} Delker, T., D. B. Pengra and P. -z. Wong,
\cprl{76}{2902}{1996}  

\bibitem[ ]{ref} Derrida, B., and H. Spohn, \jsp{51}{817}{1988}.

\bibitem[ ]{ref} Doherty, J. P.,  M. A. Moore, J. M. Kim and A. J. Bray,
\cprl {72} {2041} {1994}.

\bibitem[ ]{ref} Eden M., 1958, in {\it Symposium on information Theory
in Biology,} edited by H. P. Yockey, (Pergamon Press, New York), p. 359.

\bibitem[ ]{ref} Edwards, S. F., and D. R.  Wilkinson, 1982, Proc. R.
Soc. London A {\bf 381}, 17.

\bibitem[ ]{ref} Family, F., and T. Vicsek, 1991, {\it Dynamics of
Fractal Surfaces} (World Scientific, Singapore).

\bibitem[ ]{ref} Family, F., and T. Vicsek, \jpa{18}{L75}{1985}.

\bibitem[ ]{ref} Fisher, D. S., \cprl{56}{1964}{1986}.

\bibitem[ ]{ref} Fisher, D. S., and D. A. Huse, \cprb{43}{10728}{1991}.

\bibitem[ ]{ref} Forster, D., D. R. Nelson  and M. J. Stephen,
\cpra{16}{732}{1977}.

\bibitem[ ]{ref} Frey, E., and U. C. T\"auber, \cpre{50}{1024}{1994}.

\bibitem[ ]{ref} Galluccio, S., Y.-C. Zhang, \cpre{51}{1686}{1995}  

\bibitem[ ]{ref} Giugliarelli, G., and A. L. Stella, 1991, Physica
Scripta T{\bf 35}, 34.

\bibitem[ ]{ref} Giugliarelli, G., and A. L. Stella, 1994, Physica A
{\bf 212}, 12.

\bibitem[ ]{ref} Golubovi\'{c}, L., and R. P. U. Karunasiri,
\cprl{66}{3156}{1991}. 

\bibitem[ ]{ref} Grossmann, B., H. Guo and M. Grant,
\cpra{43}{1727}{1991}.

\bibitem[ ]{ref} Halpin-Healy, T., and Y.-C. Zhang, 1995, Phys. Rep.
{\bf 254}.

\bibitem[ ]{ref} He, S., G. L. M. K. S. Kahanda and P. -z. Wong,
\cprl{69}{3731}{1992}

\bibitem[ ]{ref} Herrasti, P., P. Oc\'on, L. Vasquez, R. C. Salvarezza,
J. M. Vara and A. J. Arvia, \cpra{45}{7440}{1992}.

\bibitem[ ]{ref} Hohenberg, P. C., and B. I. Halperin, 1977, Rev. Mod.
Phys. {\bf 49}, 435.

\bibitem[ ]{ref} Horv\'ath V. K., F. Family and T. Vicsek,
\cprl{67}{3207}{1991}.

\bibitem[ ]{ref} Hunt, A. W., C. Orme, D. R. M. Williams, B.G . Orr, and
L. M. Sander, 1994, Europhys. Lett. {\bf 27}, 611.

\bibitem[ ]{ref} Huse, D. A., J. G. Amar and F. Family,
\cpra{41}{7075}{1990}.

\bibitem[ ]{ref} Hwa, T., and M. Kardar, \cprl{62}{1813}{1989}.

\bibitem[ ]{ref} Jensen, M. H., and I. Procaccia, 1991, J. Physique II,
{\bf 1}, 1139.

\bibitem[ ]{ref} Ji, H., M. O. Robbins, \cprb{46}{14519}{1991}  

\bibitem[ ]{ref} Jiang, Z., and H. G. E. Hentschel,
\cpra{45}{4196}{1992}.

\bibitem[ ]{ref} Kapral, R., R. Livi, G. Oppo and A. Politi, \cpre {49}
{2009} {1994}.

\bibitem[ ]{ref} Kardar, M., 1994, Turkish J. of Phys. {\bf 18}, 221. 

\bibitem[ ]{ref} Kardar, M., G. Parisi and Y.-C. Zhang,
\cprl{56}{889}{1986}.

%\bibitem[ ]{ref} Kardar, M., and J. O. Indekeu, \cprl{65}{662}{1990}.

\bibitem[ ]{ref} Kardar, M., and Y.-C. Zhang, \cprl{58}{2087}{1987}.

\bibitem[ ]{ref} Keblinski, P., A. Maritan, F. Toigo and J. R. Banavar,
\cpre{49}{R4795}{1994}.

\bibitem[ ]{ref} Keblinski, P., A. Maritan, F. Toigo and J. R. Banavar,
\cprl{74}{1783}{1995}.

\bibitem[ ]{ref} Keblinski, P., A. Maritan, F. Toigo, J. Koplik and J.
R. Banavar, \cpre{49}{R937}{1994}.

\bibitem[ ]{ref} Keblinski, P., A. Maritan, F. Toigo, R. Messier and J.
R. Banavar, \cpre{53}{759}{1996}.

\bibitem[ ]{ref} Kessler, D. A., H. Levine, and L. M. Sander,
\cprl{69}{100}{1992}.

\bibitem[ ]{ref} Kim, J. M.,  and J. M. Kosterlitz,
\cprl{62}{2289}{1989}.

\bibitem[ ]{ref} Kim, J. M., M. A. Moore, and A. J. Bray,
\cpra{44}{2345}{1991}.

\bibitem[ ]{ref} Kim, J. M., and S. Das Sarma, \cprb{51}{1889}{1995}.

\bibitem[ ]{ref} Kim, J. M., and S. Das Sarma, \cprl{72}{2903}{1994}.

\bibitem[ ]{ref} Ko, D. Y. K., and  F. Seno, \cpre {50} {R1741} {1994}.

\bibitem[ ]{ref} Koplik, J.,  and Levine H., \cprb{32}{280}{1985}

\bibitem[ ]{ref} Kotrla, M., A. C. Levi, and P. Smilauer,
\euphl{20}{25}{1992}.

\bibitem[ ]{ref} Krug, J., M. Pliscke, and M. Siegert,
\cprl{70}{3271}{1993}.

\bibitem[ ]{ref} Krug, J., \jpa{22}{L769}{1989}.

\bibitem[ ]{ref} Krug, J., \jpI{1}{9}{1991}.

\bibitem[ ]{ref} Krug, J., \cprl{72}{2907}{1994}.

\bibitem[ ]{ref} Krug, J., and H. Spohn, 1990, in {\em Solids Far From
Equilibrium: Growth, Morphology and Defects, } edited by C.
Godr$\grave{\rm e}$che (Cambridge University Press), p. 479.

\bibitem[ ]{ref} Krug, J., and P. Meakin, \cpra{40}{2064}{1989}.

\bibitem[ ]{ref} Krug, J., and P. Meakin, \cpra{43}{900}{1991}.

\bibitem[ ]{ref} Lai, Z. W., and S. Das Sarma, \cprl{66}{2348}{1991}.

\bibitem[ ]{ref} Lam, C.-H., L. M. Sander, and D. E. Wolf,
\cpra{46}{R6128}{1992}.

\bibitem[ ]{ref} Lam, C.-H., and L. M. Sander, \cpre{48}{979}{1993}.

\bibitem[ ]{ref} Leschhorn, H., 1993, Physica A {\bf 200}, 136.

\bibitem[ ]{ref} Liu, D.  and M. Plischke, \cprb{38}{4781}{1988}.


\bibitem[ ]{ref} Ma, S. K., and G. F. Mazenko, \cprb{11}{4077}{1975}.

\bibitem[ ]{ref} Makse, H. A., \cpre {52} {4080} {1995}.

\bibitem[ ]{ref} Mandelbrot, B. B., 1986, in {\it Fractals in
Physics\/}, eds. L. Pietronero and E. Tosatti (Elsevier Science
Publishers) Amsterdam, pp. 3, 17, 21. 

\bibitem[ ]{ref} Maritan, A., F. Toigo, J. Koplik  and J. R. Banavar,
\cprl{69}{3193}{1992}.

\bibitem[ ]{ref} Marsili, M. and A. J. Bray, \cprl{73}{2750}{1996}.

\bibitem[ ]{ref} Marsili, M., A. Maritan, F. Toigo, and J. R. Banavar,
1996, Europhys. Lett. submitted.

\bibitem[ ]{ref} Martys, N., M. O. Robbins  and M. Cieplak,
\cprb{44}{12294}{1991}.

\bibitem[ ]{ref} Mazor, A., D. J. Srolovitz, P. S. Hagan and B. G.
Bukiet, \cprl{60}{424}{1988}.

\bibitem[ ]{ref} Meakin, P., P. Ramanlal, L. M. Sander and R. C. Ball,
\cpra{34}{5091}{1986}.

\bibitem[ ]{ref} Meakin, P., 1988 a, in {\em Phase Transition and Critical
Phenomena}, {\bf 12}, edited by C. Domb and J. L. Lebowitz
(Academic Press, London), p. 335.

\bibitem[ ]{ref} Meakin, P., \cpra{38}{994}{1988b}.

\bibitem[ ]{ref} Meakin, P., 1993, Phys. Rep. {\bf 235}, 189.

\bibitem[ ]{ref} Meakin, P., A. Coniglio, H. E. Stanley and T. A.
Witten, \cpra{34}{3325}{1986}.


\bibitem[ ]{ref} Medina, E., T. Hwa,  M. Kardar and Y.-C. Zhang,
\cpra{39}{3053}{1989}.

\bibitem[ ]{ref} Messier, R., and J. E. Yehoda, 1985, J. Appl. Phys.
{\bf 58}, 3739.

\bibitem[ ]{ref} Narayan, O., and D. S. Fisher, \cprb{48}{7030}{1993}.

\bibitem[ ]{ref} Nattermann, T., S. Stepanov, L.-H. Tang and H.
Leschhorn, 1992, J. Phys. II France {\bf 2}, 1483.

\bibitem[ ]{ref} Nattermann, T., \jpc{18}{6661}{1985}.

\bibitem[ ]{ref} Niemeyer, L., L. Pietronero and H. J. Wiesmann,
\cprl{52}{1033}{1984}.

\bibitem[ ]{ref} Nittmann, J., G. Daccord and H. E. Stanley, 1985,
Nature {\bf 314}, 141.

\bibitem[ ]{ref} Nolle, C. S., B. Koiller, N. Martys and M. O. Robbins,
\cprl{71}{2074}{1993}.

\bibitem[ ]{ref} Olami, Z., I. Procaccia and R. Zeitak,
\cpre{49}{1232}{1994}.

\bibitem[ ]{ref} Pal, S., and D. P. Landau, \prb{49}{10597}{1994}.

\bibitem[ ]{ref} Parisi G.,\euphl{17}{673}{1992}.

\bibitem[ ]{ref} Peng, C.-K., S. Havlin, M. Schwartz and H. E. Stanley,
\cpra{44}{R2239}{1991}.

\bibitem[ ]{ref} Pfeifer, P., Y. J. Wu, M. W. Cole, and J. Krim,
\cprl{62}{1997}{1989}.

\bibitem[ ]{ref} Pietronero, L., and E. Tosatti, 1985, "Fractals in
Physics" (North-Holland).

\bibitem[ ]{ref} Plischke, M., and Z. R\'acz, \cpra{32}{3825}{1985}.

\bibitem[ ]{ref} R\'acz Z., Siegert M., Liu D., and Plischke
M., \cpra {43} {5275} {1991}.

\bibitem[ ]{ref} Salvarezza, R. C., L. Vasquez, P. Herrasti, P. Oc\'on,
Vara J.  M. and Arvia A. J., \euphl{20}{727}{1992}.

\bibitem[ ] {ref} Schroeder, M., M. Siegert, D. E. Wolf, J. D. Shore
and M. Plischke, \euphl {24} {563} {1993}.

\bibitem[ ]{ref} Siegert, M., and M. Plischke, \cprl{68}{2035}{1992}.

\bibitem[ ]{ref} Siegert, M., and M. Plischke. \cprl{73}{1517}{1994}

\bibitem[ ]{ref} Sivashinsky, G. I., 1977, Acta Astron. {\bf 4}, 1177.

\bibitem[ ]{ref} Sivashinsky, G. I., 1979, Acta Astron. {\bf 6}, 560.

\bibitem[ ]{ref} Sneppen, K., \cprl{69}{3539}{1992}.

\bibitem[ ]{ref} Stokes, J. P., A. P. Kushnick and M. O. Robbins,
\cprl{60}{1386}{1988}  

\bibitem[ ]{ref} Sun, T., H. Guo  and M. Grant, \cpra{40}{6763}{1989}.

\bibitem[ ]{ref} Sun, T., and M. Plischke, \cpre {50} {3370} {1994}.

\bibitem[ ]{ref} Sun, T., and M. Plischke, \cpre{49}{5406}{1994}.

\bibitem[ ]{ref} Sun, T., and M. Plischke, \cprl{71}{3174}{1993}.

\bibitem[ ]{ref} Tang, C., S. Alexander, and R. Bruinsma,
\cprl{64}{772}{1990}.

\bibitem[ ]{ref} Tang, C., and S. Liang, \cprl {71} {2769} {1993}.

\bibitem[ ]{ref} Tang, L.-H., B. M. Forrest  and D. E. Wolf,
\cpra{45}{7162}{1992}. 

\bibitem[ ]{ref} Tang, L.-H., and H. Leschhorn, \cpra{45}{R8309}{1992}.

\bibitem[ ]{ref} Tang, L.-H., and H. Leschhorn, \cprl{70}{3832}{1993}.

\bibitem[ ]{ref} Tang, L.-H., and T. Nattermann, \cprl{66}{2899}{1991}.

\bibitem[ ]{ref} Tu, C. W., and  J. S.  Harris Jr., 1991, J. Cryst. Growth
{\bf 111}, R 9.

\bibitem[ ]{ref} Villain, J., A. Pimpinelli, and D. Wolf, 1992, Comm.
Cond. Mat. Phys. {\bf 15}, 1.

\bibitem[ ]{ref} Villain, J., \jpI{1}{19}{1991}.

\bibitem[ ]{ref} Vvedensky, D. D., A. Zangwill, C. N. Luse and M. R.
Wilby, \cpre{48}{852}{1993}.

\bibitem[ ]{ref} Witten, T. A., and L. M. Sander, \cprl{47}{1400}{1981}. 

\bibitem[ ]{ref} Wolf, D. E., \cprl{67}{1783}{1991}.

\bibitem[ ]{ref} Zhang, Y.-C.,  \jpq{51}{2129}{1990}.

\bibitem[ ]{ref} Zinn-Justin, J., 1993, {\it Quantum Field Theory and
Critical Phenomena}, (Oxford Science Publishers).

\end{thebibliography}
\end{document}